\documentclass[10pt,aps,prd,twocolumn,amsmath,amssymb,nofootinbib,eqsecnum,superscriptaddress]{revtex4-2}
\usepackage{url}
\usepackage[english]{babel}
\usepackage{graphicx}
\usepackage{amsmath,bm}
\usepackage[breaklinks,colorlinks,citecolor=blue]{hyperref}
\usepackage{hhline}
\usepackage{amssymb}
\usepackage{amsthm}
\usepackage{mathrsfs}
\usepackage{braket}
\usepackage{xcolor}
\usepackage{amssymb}
\usepackage[scr=boondox]{mathalfa}
\usepackage[normalem]{ulem}
\usepackage{tensor}
\usepackage{comment}
\usepackage{lipsum}
\usepackage{pgf}
\usepackage{tikz}
\usetikzlibrary{arrows,automata}
\usetikzlibrary{decorations.markings}
\usepackage[latin1]{inputenc}
\usetikzlibrary{positioning}
\usetikzlibrary{calc}
\usetikzlibrary{arrows.meta}

\begin{document}
\title{Quantum entanglement of masses with non-local gravitational interaction}

\author{Ulrich K. Beckering Vinckers}
\affiliation{Cosmology and Gravity Group, Department of Mathematics and Applied Mathematics,
University of Cape Town, Rondebosch 7701, Cape Town, South Africa}
\affiliation{Van Swinderen Institute, University of Groningen, 9747 AG Groningen, The Netherlands}

\author{\'{A}lvaro de la Cruz-Dombriz}
    \affiliation{ Departamento de F\'{i}sica Fundamental, Universidad de Salamanca, 37008 Salamanca, Spain}
   \affiliation{Cosmology and Gravity Group, Department of Mathematics and Applied Mathematics,
University of Cape Town, Rondebosch 7701, Cape Town, South Africa}

\author{Anupam Mazumdar}
\affiliation{Van Swinderen Institute, University of Groningen, 9747 AG Groningen, The Netherlands}
    
\begin{abstract}
We examine the quantum gravitational entanglement of two test masses in the context of linearized General Relativity with specific non-local interaction with matter.  To accomplish this, we consider an energy-momentum tensor describing two test particles of equal mass with each possessing some non-zero momentum. After discussing the quantization of the linearized theory, we compute the gravitational energy shift which is operator-valued in this case. As compared to the local gravitational interaction, we find that the change in the gravitational energy due to the self-interaction terms is finite. We then move on to study the quantum gravity induced entanglement of masses for two different scenarios. The first scenario involves treating the two test masses as harmonic oscillators with an interaction Hamiltonian given by the aforesaid gravitational energy shift. In the second scenario, each of the test masses is placed in a quantum spatial superposition of two locations, based on their respective spin states, and their entanglement being induced by the gravitational interaction and the shift in the vacuum energy. For these two scenarios, we compute both the concurrence and the von Neumann entropy; showing that an increase in the non-locality of the gravitational interaction results in a decrease in both of these quantities. 
\end{abstract}

\maketitle
\section{Introduction}

Einstein's General Relativity (GR) is successful when its predictions are compared with experiments at large distances, such as the observations from solar system tests or the detection of gravitational waves~\cite{Will:2014kxa,LIGOScientific:2016aoc}, amongst others. However, this classical theory fails at very short distances and early cosmological times, predicting for instance cosmological and black-hole singularities where the notion of space-time breaks down~\cite{ellis}. It is believed that some quantum gravity paradigm would resolve some of these questions~\cite {Kiefer:2007ria}. However, there is no laboratory proof yet of gravity being a quantum-compatible entity. In fact, we are not even aware whether gravity obeys the rules of quantum mechanics or not.

Recently, a proposal to test the quantum nature of gravity by witnessing the spin entanglement between the two quantum superposed test masses, known as quantum gravity induced entanglement of masses (QGEM), has been made \cite{Bose:2017nin,ICTS}, see also \cite{marletto2017gravitationally}. Also, alternative proposals aim to test the spin-2 nature of the gravitational interaction by witnessing the entanglement between a quantum system and a photon~\cite{Biswas:2022qto}.  For all these experimental protocols, the heart of the argument is based on  the so-called {\it Local Operation and Classical Communication} (LOCC) theorem, which states that one cannot entangle two quantum systems if they were not entangled to begin with~\cite{Bennett_1996,wooters1}. This would mean that two test masses would entangle in the presence of gravitational interaction provided gravity obeys the rules of quantum mechanics~\cite{Marshman:2019sne,Bose:2022uxe}, see also~\cite{Carney_2019,Belenchia:2018szb,Danielson:2021egj,Christodoulou:2022mkf}. As a natural consequence, a classical gravitational interaction with matter will not yield any entanglement whatsoever as shown in \cite{Marshman:2019sne,Bose:2022uxe}. 

In this context,  we are testing a quantum feature of gravity at the lowest order, which provides Newton's gravitational potential, in a similar spirit to a Bell-inequality test on quantum systems~\cite{Hensen:2015ccp}, where the correlation does not vanish even if one takes $\hbar \rightarrow 0$, as was first shown in the context of two entangled large spins~\cite{PhysRevA.46.4413,GISIN199215}. 

An interesting twist to this above discussion arises when we wish to introduce non-local interaction between matter and gravity, see~\cite{Marshman:2019sne}. Indeed, this immediately forces us to put into question one of the key assumptions behind 
LOCC, which reflects a local operation. By local, herein we mean a local unitary transformation. The question we can ask is how introducing non-local quantum gravitational interaction with matter would affect the entanglement features between two test masses.

Non-local interactions can be described by non-local field theories, see for example~\cite{Tomboulis:2015esa,Buoninfante:2018mre}, where there exists a new non-local scale, see in the context of gravity~\cite{Edholm:2016hbt}. In fact, there also exist non-local versions of gravity~\cite{Krasnikov:1987yj,Tomboulis:1997gg,Siegel:2003vt,Biswas:2005qr,Biswas,Modesto:2011kw,Deser:2007jk,Woodard:2014iga}. A very particular form of non-locality arises in string field theory~\cite{Siegel:2003vt,deLacroix:2017lif} and p-adic strings \cite{Freund:2017aqf,FREUND1987191}. In fact, in \cite{Abel:2019zou}  it was suggested  that non-local field theories may even arise naturally by discarding the higher modes of a first-quantized string theory, in a particle approximation, which promotes non-locality to the status of a fundamental scale. 

A non-local gravitational action can be recast in terms of a quadratic action of gravity, which modifies the ultraviolet (UV) behavior of gravity. The gravitational action contains infinitely many derivatives and, in order to avoid introducing ghost degrees of freedom, one takes specific analytic entire functions, which recovers the local limit of GR smoothly~\cite{Biswas,Edholm:2016hbt}. These theories can resolve the cosmological singularity, as firstly shown in~\cite{Biswas:2005qr,Biswas:2010zk,Biswas:2012bp} and the anisotropic Kasner singularity~\cite{Kumar:2021mgc}. Such theories also do not permit point-like singularities~\cite{Biswas:2005qr,Biswas,Frolov:2015bia,Frolov:2015usa,Buoninfante:2018xiw,Boos:2018bxf,Kolar:2021oba,Vinckers:2022the,Kolar:2020bpo}, and there are hints that they can even resolve astrophysical black-hole singularities~\cite{Frolov:2015bta,Koshelev:2018hpt,Buoninfante:2019swn,Buoninfante:2018rlq,Buoninfante:2018xif}. Exact solutions in the context of infinite-derivative gravity have been considered in \cite{Kilicarslan:2019njc,Dengiz:2020xbu,Kolar:2021rfl,Kolar:2021uiu} and the Hamiltonian formulation for non-local theories has been studied in \cite{llosa1994hamiltonian,Gomis:2000gy,Gomis:2003xv,Kolar:2020ezu,Heredia:2021wja,Heredia:2022mls}. It is also believed that non-local field theories may ameliorate renormalizable properties of gravity~\cite{Tomboulis:2015esa,Modesto:2017hzl,Tomboulis:1997gg,Modesto:2014lga,Modesto:2011kw,Boos:2021lsj,Boos:2021jih,Boos:2021chb}. All these effects are due to the fact that the gravitational interaction with matter weakens the UV, and this makes such theories interesting to study. In the following we shall explore the consequence of a non-local gravitational interaction in the QGEM experiment, a similar vein as testing the quantum version of the equivalence principle~\cite{Bose:2022czr}, and probing the anti-de Sitter spacetime in the context of a warped extra dimension~\cite{Elahi:2023ozf}.

 The aim of this investigation will be to study the entanglement provided a non-local gravitational interaction with matter is present. In particular, we shall study how the entanglement builds up in a momentum basis, a technique which has never been studied before in this context. Furthermore, we shall also show that the change in the gravitational energy is always finite as compared to the local gravitational interaction with matter. Consequently,  the entanglement never blows up in this class of theories, which is a novel result. To examine the entanglement, we shall focus on the concurrence as well as the von Neumann entropy, which are simple tools for estimating how the two bipartite states are entangled.

This work is organized as follows. In Section \ref{sec:IDG_review} we briefly discuss a ghost-free infinite-derivative modification of the quadratic Einstein-Hilbert (EH) action. Subsequently, we perform a redefinition of fields so that the resulting action has both local kinetic terms and a non-local interaction. Then in Section \ref{sec:quantisation} we review the quantization procedure presented in \cite{Gupta}.
In Section \ref{sec:entangle} we consider the energy-momentum tensor for a specific non-local interaction and calculate the shift in the gravitational energy. Details regarding the derivation of the gravitational energy shift are given in Appendix \ref{sec:appendix}. In Section \ref{sec:concurrence} we compute both the concurrence and von Neumann entropy for the entanglement of two test masses in the context of two different scenarios: in Subsection \ref{sec:gaussian_main} we consider the first scenario which involves treating the two test masses as harmonic oscillators and we extend the results of \cite{Bose:2022uxe} for the case of GR;  then in Subsection \ref{sec:non_gaussian_main}, the second scenario involving the spatial splitting of two test masses based on their spin \cite{Bose:2017nin} is considered in a parallel setup~\cite{Nguyen:2019huk}. In both scenarios, the entanglement is induced via the gravitational energy shift derived in Section \ref{sec:entangle}. The conclusions of this work are stated in Section \ref{sec:conclusions}.

\section{infinite-derivative modification}\label{sec:IDG_review}

In the present work, we are interested in studying linearized gravity around a Minkowski background $\eta_{\mu\nu}=\text{diag}\left(-1,1,1,1\right)$ with $\mu,\nu\in\{0,1,2,3\}$. To this end, we shall study the perturbed metric $h_{\mu\nu}$ which is related to the full metric $g_{\mu\nu}$ by:
\begin{align}
h_{\mu\nu}=\frac{1}{\kappa}\left(g_{\mu\nu}-\eta_{\mu\nu}\right)\,,
\end{align}
where  $\kappa:=\sqrt{16\pi G}=\sqrt{16\pi/M_{\text p}^2}$ and $M_{\text p}$ is the Planck mass. For the moment we make use of natural units, i.e., $c=\hbar=1$, however, we will reintroduce 
appropriate units
in the last steps when calculating the entanglement of two test masses. Let us now consider the total quadratic action of interest which in general includes three contributions:
\begin{align}\label{eq:action_total}
S=S_{\text G}+S_{\text m}+S_{\text{GF}}\,,
\end{align}
where $S_{\text G}$, $S_{\text m}$ and $S_{\text{GF}}$ are the gravitational, matter, and gauge-fixing actions, respectively. For the quadratic gravitational action, we take a special class of the most general infinite-derivative gravity theories considered in four dimensions \cite{Biswas}, namely\footnote{The quadratic action for the most general ghost-free infinite-derivative gravity theory in four-dimensions, as considered in \cite{Biswas}, admits two analytic and non-zero non-local operators. In the following, however, we shall limit our consideration to the case where we have only one such operator $\mathscr{F}\left(\Box\right)$. Nevertheless, the choice used here where only one non-zero and analytic operator is considered ensures that the infrared behaviour of the theory coincides with that of GR \cite{Biswas}.}
\begin{align}
S_{\text G}=\frac14\int\text{d}^4x\big[h^{\mu\nu}\mathscr{F}\left(\Box\right)\Box h_{\mu\nu}-h\mathscr{F}\left(\Box\right)\Box h\nonumber\\
+2h^{\mu\nu}\mathscr{F}\left(\Box\right)\partial_\mu\partial_\nu h-2h^\mu\phantom{}_\alpha\mathscr{F}\left(\Box\right)\partial_\mu\partial_\nu h^{\nu\alpha}\big]\,,
\label{SG}
\end{align}
where the $\mathscr{F}\left(\Box\right)$ may contain infinitely many derivatives and the d'Alembertian operator is given by $\Box=\partial^{\mu}\partial_{\mu}$.
The relative sign difference between the terms in \eqref{SG} above is due to the fact that such an action has to satisfy the Bianchi-identities and should also recover the usual quadratic EH action around the Minkowski background in the limits $\Box\rightarrow 0$ and $\mathscr{F}\left(\Box\right) \rightarrow 1$. In general, the higher-derivative action of gravity will be plagued by ghosts, whose degrees of freedom must be canceled. 
As noted in \cite{Biswas,Biswas:2005qr}, for any $\mathscr{F}\left(\Box\right)$, the conservation of $\delta S_{\text G}/\delta h^{\mu\nu}$ is preserved. However, in order for the theory to not admit any additional degrees of freedom compared to GR, we must require that the operator $\mathscr{F}\left(\Box\right)$ be analytic with no zeros, which will certainly constrain the form of $\mathscr{F}\left(\Box\right)$. Thus, in the following, we shall assume the non-local operator $\mathscr{F}(\Box)$ to be of the form
\begin{align}\label{eq:exp_form_factor}
\mathscr{F}\left(\Box\right)={\text e}^{-\ell^2\Box}\,,
\end{align}
where $\ell\geq0$ is referred to as the \textit{length scale of non-locality} \cite{Biswas:2005qr,Biswas}. By sending $\ell \rightarrow 0$, or $\Box \rightarrow 0$, we fully recover GR, i.e., $\mathscr{F}\left(\Box\right)\rightarrow 1$. 

For the matter action, we use
\begin{align}\label{eq:matter_action}
S_{\text m}=-\frac\kappa2\int\text{d}^4x\,h^{\mu\nu}T_{\mu\nu}\,,
\end{align}
where $T_{\mu\nu}$ is the energy-momentum tensor. 

Finally, to fix the gauge, we introduce the gauge-fixing action
\begin{align}\label{eq:gauge_fix_action}
S_{\text{GF}}=-\frac12\int\text{d}^4x&\left(\partial_\mu h^\mu\phantom{}_\nu-\frac12\partial_\nu h\right)\nonumber\\
&\times\mathscr{F}\left(\Box\right)\left(\partial_\alpha h^{\alpha\nu}-\frac12\partial^\nu h\right)\,,
\end{align}
which specifies the so-called Harmonic or de Donder gauge condition. Substituting equations~\eqref{SG}, \eqref{eq:matter_action} and~\eqref{eq:gauge_fix_action} into the total action \eqref{eq:action_total} gives~\footnote{While there are contributions arising from total derivatives, we shall ignore them in our analysis.} 
\begin{align}\label{eq:action_total_2}
S=\frac14\int\text{d}^4x\bigg[h^{\mu\nu}\mathscr{F}\left(\Box\right)\Box\left(h_{\mu\nu}-\frac12h\eta_{\mu\nu}\right)-2\kappa h^{\mu\nu}T_{\mu\nu}\bigg]\,.
\end{align}
By introducing the redefined field
\begin{align}
\gamma_{\mu\nu}:=\mathscr{F}^{1/2}\left(\Box\right)\left(h_{\mu\nu}-\frac12h\eta_{\mu\nu}\right)\,,
\end{align}
which coincides with the redefined field used in \cite{Gupta,Einstein:1918btx} in the local case $\mathscr{F}=1$, we can write the total action \eqref{eq:action_total_2} as
\begin{align}\label{eq:action_with_interaction_redef} 
S=-\frac14&\int\text{d}^4x\bigg[\partial_\alpha\gamma^{\mu\nu}\partial^\alpha\gamma_{\mu\nu}-\frac12\partial^\mu\gamma\partial_\mu\gamma\nonumber\\
&+2\kappa\left(\gamma^{\mu\nu}-\frac12\eta^{\mu\nu}\gamma\right)\mathscr{F}^{-1/2}\left(\Box\right)T_{\mu\nu}\bigg]\,.
\end{align}
The action \eqref{eq:action_with_interaction_redef} now contains local kinetic terms while possessing a non-local interaction term. In the following section, we shall review Gupta's quantization procedure for such a theory.

\section{Quantization of the linearized gravity}\label{sec:quantisation}

Following \cite{Gupta}, let us consider the action defined as in \eqref{eq:action_with_interaction_redef} but with $\gamma_{\mu\nu}$ and $\gamma$ being treated as independent fields. In addition, we consider the case of a vanishing energy-momentum tensor, i.e., we set $T_{\mu\nu}=0$, and perform the quantization procedure as in \cite{Gupta}. Accordingly, $\gamma_{\mu\nu}$ and $\gamma$  collectively contain now eleven components. Furthermore, $\gamma_{\mu\nu}$ and $\gamma$ each have a canonical conjugate momentum as well as their own set of commutation relations. Thus, by introducing the canonical momenta and imposing the usual commutation relations, one can show that \cite{Gupta}
\begin{align}
\left[\gamma_{\mu\nu}(x),\gamma_{\alpha\beta}(x')\right]&=i\left(\eta_{\mu\alpha}\eta_{\nu\beta}+\eta_{\mu\beta}\eta_{\nu\alpha}\right)D\left(x-x'\right)\,,\label{eq:commutation_relation_combined}\\
\left[\gamma(x),\gamma(x')\right]&=-4iD(x-x')\,,\label{eq:commutation_schwinger_gamma}
\end{align}
having followed Schwinger's notation \cite{Schwinger:1948yk}. Let us now expand the $\gamma_{\mu\nu}$ and $\gamma$ fields in terms of Fourier modes, yielding
\begin{align}\label{eq:gamma_mu_nu_Fourier}
\gamma_{\mu\nu}=\frac{1}{(2\pi)^{3/2}}\int\frac{\text{d}^3k}{\sqrt{2\omega_{\bm k}}}\left[a_{\mu\nu}\left(\bm k\right){\text e}^{ikx}+a^\dagger_{\mu\nu}\left(\bm k\right){\text e}^{-ikx}\right]\,,
\end{align}
and
\begin{align}\label{eq:gamma_Fourier}
\gamma=\frac{2}{(2\pi)^{3/2}}\int\frac{\text{d}^3k}{\sqrt{2\omega_{\bm k}}}\left[b\left(\bm k\right){\text e}^{ikx}+b^\dagger\left(\bm k\right){\text e}^{-ikx}\right]\,,
\end{align}
respectively, and where 
we have used $k^0=\omega_{\bm k}=|\bm k|$. 
At this point we note that $\gamma_{\mu\nu}\rightarrow \hat\gamma_{\mu\nu}$ and 
similarly $\gamma \rightarrow \hat\gamma$, i.e., these fields are now treated as quantum operators. Analogously, $h_{\mu\nu}\rightarrow \hat h_{\mu\nu}$ and $T_{\mu\nu} \rightarrow \hat T_{\mu\nu}$. However, to avoid any cluttering of the upcoming formulae, we will not explicitly write $\hat{}$ on top of the operators, but it is assumed everywhere from now on-wards that positions, momenta, and gravitational degrees of freedom are all quantum operators.

By making use of the commutation relations \eqref{eq:commutation_relation_combined} and \eqref{eq:commutation_schwinger_gamma}, one can obtain the following commutation relations for the Fourier modes
\begin{align}
\left[a_{\mu\nu}(\bm k),a^\dagger_{\alpha\beta}(\bm k')\right]&=\left(\eta_{\mu\alpha}\eta_{\nu\beta}+\eta_{\mu\beta}\eta_{\nu\alpha}\right)\delta^{(3)}(\bm k-\bm k')\,,\label{eq:commutation_a_mu_nu}\\
\left[b(\bm k),b^\dagger(\bm k')\right]&=-\delta^{(3)}\left(\bm k-\bm k'\right)\,.\label{eq:commutation_a}
\end{align}
Finally, in terms of the Fourier modes, the Hamiltonian of the vacuum system is given by \cite{Gupta}
\begin{align}\label{eq:vacuum_hamiltonian}
H_0=\int\text{d}^3k\omega_{\bm k}\left[\frac12a^\dagger_{\mu\nu}(\bm k)a^{\mu\nu}(\bm k)-b^\dagger(\bm k)b(\bm k)\right]\,.
\end{align}
We note that there is a negative sign appearing in the right-hand sides of the commutation relations \eqref{eq:commutation_a_mu_nu} and \eqref{eq:commutation_a} for $a_{0i}\left(\bm k\right)$ and $b\left(\bm k\right)$ respectively where $i\in\{1,2,3\}$. It follows that the operators $\int\text{d}^3k\omega_{\bm k}a_{0i}^\dagger\left(\bm k\right)a_{0i}\left(\bm k\right)$ and $\int\text{d}^3k\omega_{\bm k}b^\dagger\left(\bm k\right)b\left(\bm k\right)$ have nonpositive eigenvalues \cite{Gupta,Suraj_N_Gupta_1950}. Therefore, when acting the Hamiltonian on some state, the terms containing $a_{0i}\left(\bm k\right)$ and $b\left(\bm k\right)$ operators contribute nonnegative values to the energy since their coefficients are negative in the expression \eqref{eq:vacuum_hamiltonian}. It follows that the energy values associated with the Hamiltonian \eqref{eq:vacuum_hamiltonian} acting on some general state are nonnegative. Nevertheless, there is still the issue of whether these states will have negative probabilities. It is possible to impose some supplementary conditions \cite{Gupta,Suraj_N_Gupta_1950} that result in physical states having only two polarizations and positive probabilities; thus ensuring that the Hamiltonian is bounded from below. These supplementary conditions are discussed explicitly in Appendix \ref{sec:supplementary_conditions} and we show, following \cite{Gupta,Suraj_N_Gupta_1950}, that these lead to the Hamiltonian being bounded from below.

\section{Shift in the Gravitational energy}\label{sec:entangle}

Having assumed a vanishing energy-momentum tensor, the system in Section \ref{sec:quantisation}
was described by the Hamiltonian \eqref{eq:vacuum_hamiltonian}. In order to study the QGEM, we shall now consider an energy-momentum tensor of the form
\begin{align}\label{eq:energy_momentum_with_p}
T_{\mu\nu}=\frac{ p_\mu p_\nu}{ E}\left[\delta^{(3)}\left({\bm r}-{\bm r}_A\right)+\delta^{(3)}\left({\bm r}-{\bm r}_B\right)\right]\,,
\end{align}
where $p_\mu=\left(- E,{\bm p}\right)$ and $E=\sqrt{\bm p^2+m^2}$ \cite{Bose:2022uxe}. The energy-momentum tensor \eqref{eq:energy_momentum_with_p} describes two test masses, denoted by $A$ and $B$, with some momentum. Here, we shall confine the motion of the two test masses $A$ and $B$ to the $z$-axis, so $\bm r_A=(0,0,x_A)$ and $\bm r_B=(0,0,x_B)$. We note that here $T_{\mu\nu}$ is treated as an operator according to the Weyl quantization procedure \cite{Weyl:1927vd}, i.e., all products of position and momentum operators on the right-hand side of \eqref{eq:energy_momentum_with_p} represent their symmetrization. The total Hamiltonian can be computed as follows
\begin{align}
H=H_0+\kappa V\,,
\end{align}
where, from equation \eqref{eq:action_with_interaction_redef}, the interaction Hamiltonian $V$ is
\begin{align}\label{eq:V_interaction}
V=\frac{1}2\int\text{d}^3r\left(\gamma^{\mu\nu}-\frac12\eta^{\mu\nu}\gamma\right)\mathscr{F}^{-1/2}\left(\Box\right)T_{\mu\nu}\,.
\end{align}
We denote the ground state of the vacuum system as $\ket{0}$ and define the following one-particle relativistically normalized states
\begin{align}
\ket{\bm k}_{\mu\nu}:=\sqrt{2\omega_{\bm k}}\left[a^\dagger_{\mu\nu}(\bm k)-\eta_{\mu\nu}b^\dagger(\bm k)\right]\ket{0}\,.
\end{align}
We can now make use of well-known perturbation theory (see for example \cite{michael1979advanced}) to calculate the gravitational energy shift $\Delta H$ to second order in $\kappa$
\begin{align}\label{eq:ham_trajectories}
\Delta H=-\frac{\kappa^2}{2}\int\text{d}^3k\frac{\bra{0}V\ket{\bm k}_{\mu\nu}\eta^{\mu\alpha}\eta^{\nu\beta}\tensor[_{\alpha\beta}]{\bra{\bm k}}{}V\ket{0}}{\bm k^2}\,.
\end{align}
As noted in \cite{Bose:2022uxe}, the first order ${\cal O}(\kappa)$ correction is zero since it will involve inner products of the ground state with the first excited state only. 

By substituting equations \eqref{eq:exp_form_factor} and \eqref{eq:energy_momentum_with_p} into \eqref{eq:V_interaction}, \eqref{eq:ham_trajectories} becomes
\begin{align}\label{eq:hammy_integral_non_static_to_calculate}
&\Delta H=-\frac{\kappa^2}{8}\int\text{d}^3k\left[\frac{{\mathcal T}_{00}^{\dagger}(\bm k){\mathcal T}_{00}(\bm k)}{\bm k^2}+\frac{{\mathcal T}_{33}^{\dagger}(\bm k){\mathcal T}_{33}(\bm k)}{\bm k^2}\right]\nonumber\\
&-\frac{\kappa^2}{8}\int\text{d}^3k\left[\frac{{\mathcal T}_{00}^{\dagger}(\bm k){\mathcal T}_{33}(\bm k)}{\bm k^2}+\frac{{\mathcal T}_{33}^{\dagger}(\bm k){\mathcal T}_{00}(\bm k)}{\bm k^2}\right]\nonumber\\
&+\frac{\kappa^2}{2}\int\text{d}^3k\ \frac{{\mathcal T}_{03}^{\dagger}(\bm k){\mathcal T}_{03}(\bm k)}{\bm k^2}\,,
\end{align}
where we define
\begin{align}\label{eq:Fourier_comp_T_mu_nu}
\mathcal{T}_{\mu\nu}\left(\bm k\right):&=\frac{{\text e}^{-\ell^2\bm k^2/2-i\bm k\cdot\bm r_A}}{\left(2\pi\right)^{3/2}}
\begin{pmatrix}
E_A & 0 & 0 & p_A\\
0 & 0 & 0 & 0 \\
0 & 0 & 0 & 0 \\
p_A & 0 & 0 & p_A^2/E_A
\end{pmatrix}
\nonumber\\
&+\left(A\leftrightarrow B\right)\,,
\end{align}
as the Fourier transform of ${\text e}^{\ell^2\Delta/2}T_{\mu\nu}$. We also note that $p_A$ and $p_B$ denote the $z$-components of $\bm p_A$ and $\bm p_B$ respectively. As mentioned above, $\bm r_A$, $\bm r_B$, $\bm p_A$ and $\bm p_B$ are being treated as operators with all products of position and momentum operators representing, implicitly, the symmetrization $\left\{\bm r_A,\bm p_A\right\}/2$ and similar. The bottom line is that if gravity is quantum in nature, so will the change in the gravitational energy, and consequently, the change in the gravitational energy would not be a C number, but an operator-valued quantity.


As remarked in \cite{Bose:2022uxe}, the computation of the gravitational energy shift includes both a contribution from the self-energy of the individual particles and a contribution from the interaction. To this end, let us write
\begin{align}\label{eq:H_split}
\Delta H=\Delta H_{\text{SE}}+\Delta H_{\text D}\,,
\end{align}
where $\Delta H_{\text{SE}}$ is the self-energy contribution and $\Delta H_{\text D}$ is simply defined through the $\Delta H-\Delta H_{\text{SE}}$ difference. By substituting the components \eqref{eq:Fourier_comp_T_mu_nu} into \eqref{eq:hammy_integral_non_static_to_calculate} and evaluating the integral, we find the following expression for the self-energy 
\begin{align}\label{eq:non_static_self_energy}
\Delta H_{\text{SE}}=-\frac{1}{2\ell M_{\text p}^2\sqrt\pi}\left(E_A^2+\frac{p_A^4}{ E_A^2}-2p_A^2\right)+\left(A\leftrightarrow B\right)\,.
\end{align}
It is evident from \eqref{eq:non_static_self_energy} that the self-energy is finite for $\ell>0$. We also note that the self-energy above is operator-valued, i.e., it depends on the operators $p_A$ and $p_B$. In the static case, the self-energy reduces to
\begin{align}
\lim_{p_A,p_B\rightarrow0}\Delta H_{\text{SE}}=-\frac{m^2}{\ell M_{\text p}^2\sqrt\pi}\,,
\end{align}
which is a constant. This result is in stark contrast with the local scenario, since when $\ell\rightarrow 0$, the self-energy contribution blows away at order ${\cal O}(\kappa^2)$.

Turning our attention to finding the quantity $\Delta H_{\text D}$ which describes the interaction, we find
\begin{align}\label{eq:non_static_H_D}
&\Delta H_{\text D}=-\frac{1}{M_{\text p}^2|{\bm r}_A-{\bm r}_B|}\bigg[ E_A E_B+\frac{ p_A^2 p_B^2}{ E_A E_B}\nonumber\\
&+\left(\frac{ E_A p_B^2}{ E_B}+\frac{ E_B p_A^2}{ E_A}\right)-4 p_A p_B
\bigg]\text{erf}\left(\frac{|{\bm r}_A-{\bm r}_B|}{2\ell}\right)\,.
\end{align}
For a derivation of equations \eqref{eq:non_static_self_energy} and \eqref{eq:non_static_H_D}, we direct the interested reader to Appendix \ref{sec:appendix}. As was done in \cite{Bose:2022uxe}, we expand our result up to fourth-order in the momentum operators, yielding
\begin{align}\label{eq:H_g_non_static_IDG}
&\Delta H_{\text D}\approx-\frac{1}{M_{\text p}^2|{\bm r}_A-{\bm r}_B|}\bigg[m^2+\frac{3 p_A^2-8 p_A p_B+3 p_B^2}{2}\nonumber\\
&-\frac{5 p_A^4-18 p_A^2 p_B^2+5 p_B^4}{8m^2}\bigg]\text{erf}\left(\frac{|{\bm r}_A-{\bm r}_B|}{2\ell}\right)\,.
\end{align}
We also note that taking the static limit of the last expression \eqref{eq:H_g_non_static_IDG} gives us
\begin{align}\label{eq:static_interaction_energy_D}
\lim_{p_A,p_B\rightarrow0}\Delta H_{\text D}=-\frac{m^2}{M_{\text p}^2|\bm r_A-\bm r_B|}\text{erf}\left(\frac{|\bm r_A-\bm r_B|}{2\ell}\right)\,.
\end{align}
Equation \eqref{eq:static_interaction_energy_D} provides us with the operator-valued version of the potential derived in \cite{Biswas}.

\section{Concurrence and entropy for the entanglement}\label{sec:concurrence}
\subsection{Two harmonic oscillators}\label{sec:gaussian_main}
Let us now treat the two test masses as harmonic oscillators; both with frequency $\omega_m$. Let us denote the annihilation (creation) operators for the test masses $A$ and $B$ as $a$ and $b$ ($a^\dagger$ and $b^\dagger$) respectively. 
For the position operators $x_A$ and $x_B$, we write
\begin{align}
 x_A=-\frac{d}2+\delta  x_A\,,\ \ \ \ \ \ \ \ x_B=\frac{d}2+\delta  x_B\,,
\end{align}
hence the two harmonic oscillators are placed a distance $d$ apart and are subject to small fluctuations described respectively by $\delta x_A$ and $\delta x_B$. By performing a Taylor expansion of the first term on the right-hand side of equation \eqref{eq:H_g_non_static_IDG} up to order $\left(\delta x_A-\delta x_B\right)^2$, we can extract the lowest order matter-matter interaction energy
\begin{align}\label{eq:H_AB_IDG_non_static}
&H_{\text{AB}}:=\frac{2m^2}{M_{\text p}^2d^2}\left[\frac{1}{d}\text{erf}\left(\frac{d}{2\ell}\right)-\frac{{\text e}^{-d^2/4\ell^2}}{\ell\sqrt\pi}-\frac{d^2{\text e}^{-d^2/4\ell^2}}{4\ell^3\sqrt\pi}\right]\nonumber\\
&\times\delta  x_A\delta  x_B+\left(\frac{4p_A p_B}{M_{\text p}^2d}-\frac{9p_A^2 p_B^2}{4M_{\text p}^2m^2d}\right)\text{erf}\left(\frac{d}{2\ell}\right)\,.
\end{align}
In terms of the creation and annihilation operators, the position operators take the usual form \cite{sakurai}
\begin{align}\label{eq:position_to_creation}
\delta  x_A=\frac{1}{\sqrt{2m\omega_m}}\left( a+ a\phantom{}^\dagger\right),\ \delta  x_B=\frac{1}{\sqrt{2m\omega_m}}\left( b+ b\phantom{}^\dagger\right)\,,
\end{align}
whereas for the momentum operators we have
\begin{align}
p_A=i\sqrt{\frac{m\omega_m}{2}}\left(a^\dagger-a\right)\,,\ \ \ \ p_B=i\sqrt{\frac{m\omega_m}{2}}\left(b^\dagger-b\right)\,.
\end{align}
In terms of the creation and annihilation operators $a^\dagger$, $a$, $b^\dagger$ and $b$, the Hamiltonian for the system is of second order in $\kappa$ and fourth order in the momentum expansion
\begin{align}\label{eq:hammy_HT}
&H_{\text{HO}}=\hbar\bigg[\omega_m\left(a^\dagger a+b^\dagger b\right)+\mathcal{G}_1\left( a b+ a\phantom{}^\dagger b+ a b\phantom{}^\dagger+ a\phantom{}^\dagger b\phantom{}^\dagger\right)\nonumber\\
&+\mathcal{G}_2\left( a b-a\phantom{}^\dagger b-a b\phantom{}^\dagger+ a\phantom{}^\dagger b\phantom{}^\dagger\right)+\mathcal{G}_3\left(a^\dagger-a\right)^2\left(b^\dagger-b\right)^2\bigg]\,,
\end{align}
where the functions
\begin{align}\label{eq:math_cal_G}
\mathcal{G}_1&:=\frac{mG}{d^3\omega_m}\Bigg[\text{erf}\left(\frac{d}{2\ell}\right)-\frac{d{\text e}^{-d^2/4\ell^2}}{\ell\sqrt\pi}-\frac{d^3{\text e}^{-d^2/4\ell^2}}{4\ell^3\sqrt\pi}\Bigg]\,,\\
\mathcal{G}_2&:=-\frac{2mG\omega_m}{c^2d}\text{erf}\left(\frac{d}{2\ell}\right)\,,\label{eq:mathcal_G_2}\\
\mathcal{G}_3&:=-\frac{9\omega_m^2G\hbar}{16c^4d}\text{erf}\left(\frac{d}{2\ell}\right)\,,\label{eq:mathcal_G_3}
\end{align}
have been defined. It is at this point that 
dimensionality of $\hbar$ and $c$ is reinstated.

Let us denote $\ket{0}_A$ and $\ket{0}_B$ as the ground states associated with $H_{\text{HO}}$ when there is no interaction, i.e., for $\mathcal{G}_1=\mathcal{G}_2=\mathcal{G}_3=0$. If  instead these functions are non-zero, we denote the eigenket obtained through first-order perturbation theory as $\ket{\psi}$. Using the standard perturbation theory of quantum mechanics (see \cite{sakurai} for details), we have
\begin{align}\label{eq:psi_first_order_11}
\ket{\psi}=\frac{1}{\sqrt{1+\left[\left(\mathcal{G}_1+\mathcal{G}_2\right)^2+\mathcal{G}_3^2\right]/\left(4\omega^2_m\right)}}\bigg[\ket{0}_{A}\otimes\ket{0}_{B}\nonumber\\
-\frac{\left(\mathcal{G}_1+\mathcal{G}_2\right)}{2\omega_m}\ket{1}_{A}\otimes\ket{1}_{B}-\frac{\mathcal{G}_3}{2\omega_m}\ket{2}_{A}\otimes\ket{2}_{B}\bigg]\,.
\end{align}
The density matrix associated with the test mass denoted by $A$ can be computed as follows \cite{wooters1}
\begin{align}\label{eq:rho_from_psi}
\rho_{A}=\sum_{n}\tensor[_{B}]{\braket{n|\psi}}{}\braket{\psi|n}_{B}\,.
\end{align}
That is, one first computes $\ket{\psi}\bra{\psi}$ and then traces over the $B$ eigenstates. Using the density matrix above, one can compute the concurrence $C$ through \cite{rungta}
\begin{align}\label{eq:concurrence_to_trace}
C^2=2\left[1-\text{tr}\left(\rho^2_A\right)\right]\,.
\end{align}
It follows that the concurrence is given by
\begin{align}\label{eq:concurrence_from_G}
C=\sqrt{2\left\{1-\frac{1+\left(\left(\mathcal{G}_1+\mathcal{G}_2\right)^4+\mathcal{G}_3^4\right)/\left(16\omega_m^4\right)}{\left[1+\left(\left(\mathcal{G}_1+\mathcal{G}_2\right)^2+\mathcal{G}_3^2\right)/\left(4\omega^2_m\right)\right]^2}\right\}}\,.
\end{align}

Let us now examine the effect of non-locality on the concurrence. For our specific choice of parameters, we take the harmonic oscillator frequency to be $\omega_m=2\pi\text{Hz}$ and let $A$ and $B$ be two mesoscopic masses with $m=10^{-14}\text{kg}$. Motivated by \cite{Bose:2017nin}, the minimum separation\footnote{As pointed out in \cite{Bose:2017nin}, the Casimir-Polder interaction \cite{Casimir:1947kzi,Casimir:1948dh} is a tenth of that of the Newtonian gravitational interaction at that distance.} that we shall consider is $d\sim200\mu\text{m}$. For the length scale of non-locality, we consider values of $\ell$ consistent with \cite{Edholm:2016hbt}. For such a choice of parameters, we have $\mathcal{G}_1/\left(2\omega_m\right)\approx1.06\times10^{-15}$, $\mathcal{G}_2/\left(2\omega_m\right)\approx-3.71\times10^{-38}$ and $\mathcal{G}_3/\left(2\omega_m\right)\approx-7.69\times10^{-75}$ in the local ($\ell=0$) case. The concurrence \eqref{eq:concurrence_from_G} may now be approximated as
\begin{align}\label{eq:concurrence_approx}
C\approx\frac{|\mathcal{G}_1|}{\omega_m}\,.
\end{align}

\begin{figure}
  \includegraphics[width=1\linewidth]{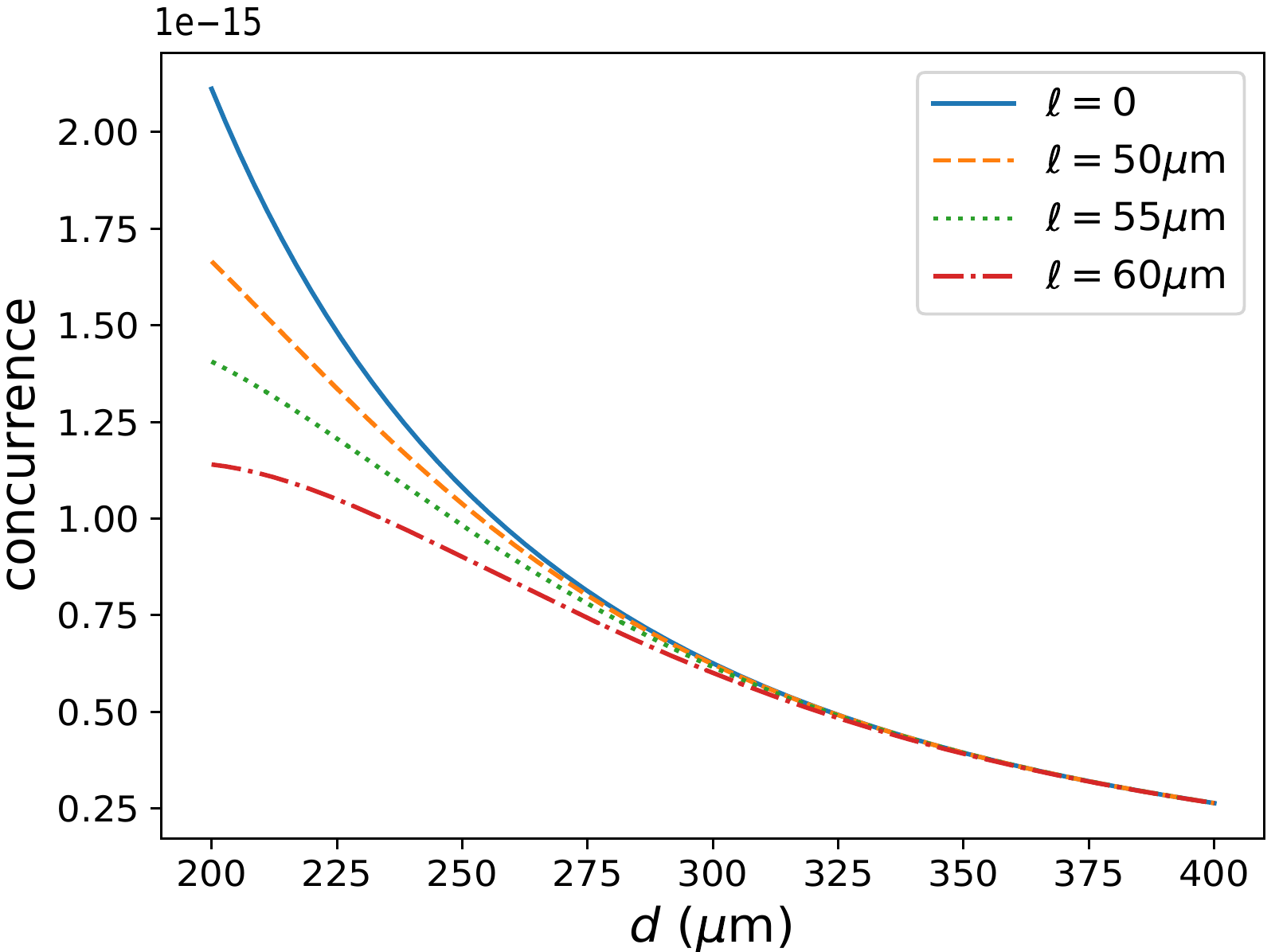}
\caption{Plots of the concurrence \eqref{eq:concurrence_from_G} using the approximation \eqref{eq:concurrence_approx}. To produce each of the profiles, we have set $m=10^{-14}\text{kg}$ and $\omega_m=2\pi$Hz. The solid blue curve corresponds to the local ($\ell=0$) case while the dashed orange, dotted green and dash-dotted red curves correspond to the cases where $\ell=50$, $55$ and $60\mu\text{m}$ respectively}
\label{fig:static_concurrence}
\end{figure}

 In Figure \ref{fig:static_concurrence} we plot the concurrence for various values of $\ell$ using the approximation \eqref{eq:concurrence_approx}. It is clear from this figure that the concurrence approaches that of the local case as the length scale of non-locality decreases. In addition, we note that all curves start to coincide as the separation $d$ increases. We also note that for smaller values of $d$ the concurrence decreases for increasing $\ell$. Lastly, for small $d/\ell$, i.e., deep in the UV or alternatively at scales where the non-local interaction is to be proved, the concurrence starts to saturate the quantum entanglement between the two harmonic oscillators. 

We note that the concurrence is of the order $10^{-15}$, i.e., beyond any reach of detectability through an experiment, and not even via tomography techniques discussed in~\cite{Barker:2022mdz}. 
Alternatively, an examination of the von Neumann entropy defined as \cite{nielsen_chuang_2010}
\begin{align}\label{eq:von_Neumann_entropy_def}
S_A:=-\text{tr}\left(\rho_A\log_2\rho_A\right)\,,
\end{align}
will also give small values. For instance, for a separation of $d=200\mu\text{m}$, we have $S_A\approx1.11\times10^{-28}$ for the local ($\ell=0$) case while $S_A\approx0.69\times10^{-28}$ for $\ell=50\mu\text{m}$. 
 

In the following subsection, we shall consider a non-Gaussian scenario for which a difference in the entanglement between local and non-local gravitational interactions may be realized experimentally in the future.

\subsection{Spatial superposition}\label{sec:non_gaussian_main}

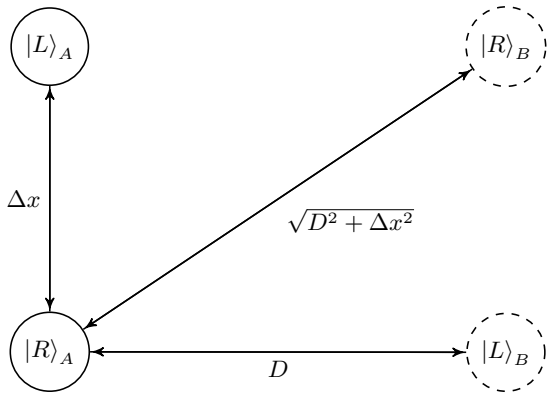
\begin{figure}
\begin{tikzpicture}[->,>=stealth',shorten >=1pt,auto,node distance=2cm,
                    semithick,decoration={
    markings,
    mark=at position 0.5 with {\arrow{>}}}]

  \node[state] (A)                    {$\ket{L}_A$};
  \node[state,dashed]         (D) [right = 5cm of A] {$\ket{R}_B$};
  \node[state]         (C) [below = 3cm of A] {$\ket{R}_A$};
  \node[state,dashed]         (F) [below = 3cm of D]       {$\ket{L}_B$};

  \path (A) edge              node [swap] {$\Delta x$} (C)
        (C) edge              node {} (A)
        (C) edge              node [swap] {$D$} (F)
        (F) edge              node {} (C)
        (C) edge              node [swap] {$\sqrt{D^2+\Delta x^2}$} (D)
        (D) edge              node {} (C);
\end{tikzpicture}
\caption{Spatial splitting of the two test masses $A$ and $B$ in the parallel QGEM setup presented in \cite{Nguyen:2019huk}. The states $\ket{L}_A$ and $\ket{R}_B$ as well as $\ket{R}_A$ and $\ket{L}_B$ are separated by a distance $D$. In addition, $\Delta x$ denotes the separation between $\ket{L}_A$ and $\ket{R}_A$ as well as between $\ket{L}_B$ and $\ket{R}_B$.}
\label{fig:diagram_SG}
\end{figure}

Here, we shall consider the parallel setup first presented in \cite{Nguyen:2019huk} and further studied in \cite{Tilly:2021qef,Schut:2021svd,Barker:2022mdz}, although herein generalized for the case where the gravitational interaction with matter is non-local. Thus, we assume that the two test masses $A$ and $B$ are each endowed with two spin states: $\ket{\uparrow}_A$, $\ket{\downarrow}_A$ and $\ket{\uparrow}_B$, $\ket{\downarrow}_B$, respectively, and separated by a spatial superposition of $\Delta x$. In addition, we take the two masses to be separated by a distance $D$. According to the experimental protocol outlined in \cite{Bose:2017nin}, two Stern-Gerlacht interferometers are used to perform a spatial splitting of the two test masses based on their spin states. Let us discuss this protocol and calculate the entanglement for the parallel setup of \cite{Nguyen:2019huk}. We start off with the state~\cite{Bose:2017nin},
\begin{align}
\ket{\psi_i}=\frac12\ket{\mathcal{C}}_A\ket{\mathcal{C}}_B\left(\ket{\uparrow}_A+\ket{\downarrow}_A\right)\left(\ket{\uparrow}_B+\ket{\downarrow}_B\right)\,,
\end{align}
where the test masses $A$ and $B$ are initially localised according to the states $\ket{\mathcal{C}}_A$ and $\ket{\mathcal{C}}_B$, respectively, and separated by a distance $D$. By performing a spatial splitting of the two test masses $A$ and $B$ through $\ket{\mathcal{C},\uparrow}_j\rightarrow\ket{L,\uparrow}_j$ and $\ket{\mathcal{C},\downarrow}_j\rightarrow\ket{R,\downarrow}_j$ for $j\in\{A,B\}$ we have at $t=0$~\cite{Bose:2017nin}
\begin{align}\label{eq:spatially_split_state}
\ket{\psi_i}\rightarrow\frac12\left(\ket{L,\uparrow}_A+\ket{R,\downarrow}_A\right)\left(\ket{L,\uparrow}_B+\ket{R,\downarrow}_B\right)\,,
\end{align}
The spatial splitting is carried out in such a way that $\ket{L}_A$ and $\ket{R}_B$, as well as $\ket{R}_A$ and $\ket{L}_B$, are separated by a distance $D$.  In Figure \ref{fig:diagram_SG}, we show the spatial splitting of the two test masses and the separation between each of the states.

The spatially-split state on the right-hand side of \eqref{eq:spatially_split_state} is allowed to evolve for a time $\tau$, being the evolution described by the operator ${\text e}^{-i\Delta H_{\text D}\tau/\hbar}$ where $\Delta H_{\text D}$ is the gravitational energy shift, which in the non-relativistic case is given by the right-hand side of \eqref{eq:static_interaction_energy_D}.  Thus, one obtains the final evolution of the state at time $t=\tau$~\cite{Bose:2017nin}:
\begin{align}\label{eq:parallel_final_state}
\ket{\psi_f}=\frac12\ket{\mathcal{C}}_A\ket{\mathcal{C}}_B\bigg[\ket{\uparrow}_A\left(\ket{\uparrow}_B+{\text e}^{i\left(\theta-\phi\right)}\ket{\downarrow}_B\right)\nonumber\\
+\ket{\downarrow}_A\left(\ket{\downarrow}_B+{\text e}^{i\left(\theta-\phi\right)}\ket{\uparrow}_B\right)\bigg]\,,
\end{align}
where the final state is an entangled state and the entanglement phase is given by:
\begin{align}
\phi&:=\frac{Gm^2\tau}{\hbar\sqrt{D^2+\Delta x^2}}\text{erf}\left(\frac{\sqrt{D^2+\Delta x^2}}{2\ell}\right)\,,\\
\theta&:=\frac{Gm^2\tau}{\hbar D}\text{erf}\left(\frac{D}{2\ell}\right)\,.
\end{align}
In the expression \eqref{eq:parallel_final_state}, we have ignored an overall phase factor of ${\text e}^{i\phi}$ without any loss of generality. Since the test masses are now once again localized in position, we can fully characterize the entanglement using the spin states, see~\cite{Bose:2017nin} for further details. Thus, we can ignore the $\ket{\mathcal{C}}_A$ and $\ket{\mathcal{C}}_B$ states in \eqref{eq:parallel_final_state} and compute the density matrix associated with $A$ as
\begin{align}\label{eq:density_matrix_non_gaussian}
\rho_A=\frac12
\begin{pmatrix}
1 & \cos\left(\theta-\phi\right)\\
\cos\left(\theta-\phi\right) & 1
\end{pmatrix}\,.
\end{align}

\begin{figure}
  \includegraphics[width=1\linewidth]{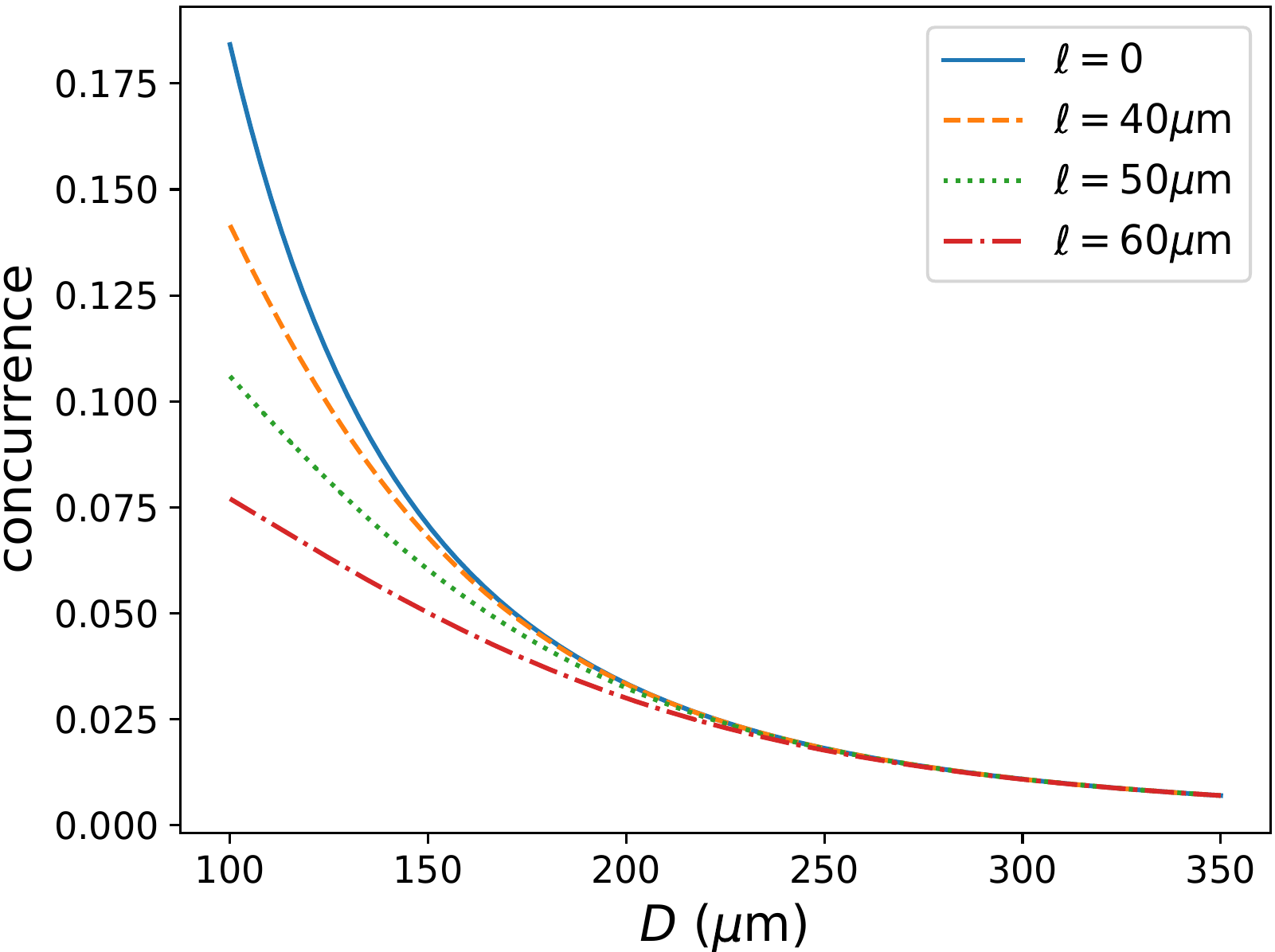}
\caption{Plots of the concurrence \eqref{eq:concurrence_non_gaussian} for the parallel set-up presented in \cite{Nguyen:2019huk} for the case of non-local gravitational interaction; describing two mesoscopic test masses, separated by a distance $D$, and spatially split based on their intrinsic spins. To produce each of the profiles, we have set $m=10^{-14}\text{kg}$. The solid blue curve corresponds to the local ($\ell=0$) case while the dashed orange, dotted green, and dash-dotted red curves correspond to the cases where $\ell=40$, $50$ and $60\mu\text{m}$ respectively}
\label{fig:static_concurrence_non_Gaussian}
\end{figure}

Having obtained the density matrix, we can compute the concurrence
\begin{align}\label{eq:concurrence_non_gaussian}
C=|\sin\left(\theta-\phi\right)|\,,
\end{align}
as well as the von Neumann entropy
\begin{align}
S_A&=-\frac12\log_2\left(\frac{\sin^2\left(\theta-\phi\right)}{4}\right)\nonumber\\
&-\frac12\cos\left(\theta-\phi\right)\log_2\left(\frac{1+\cos\left(\theta-\phi\right)}{1-\cos\left(\theta-\phi\right)}\right)\,,
\end{align}
for the entanglement. For the parameter specifications $\tau=1\,\text s$, $\Delta x=100\,\mu\text m$ and $D=200\,\mu\text m$, we have plotted the concurrence in Figure \ref{fig:static_concurrence_non_Gaussian} for different values of $\ell$. This figure shows how the concurrence grows slowly with decreasing $D/\ell$, a similar behavior as noted in Subsection \ref{sec:gaussian_main} for the case of two harmonic oscillators. For instance, at $D=200\,\mu\text m$, the concurrence in the GR case is $C\approx3.34\times10^{-2}$, while for the $\ell=50\,\mu\text m$ case, the concurrence is $C\approx3.23\times10^{-2}$. In addition, for the same parameter specifications, we note that the von Neumann entropy  for the GR case is $S_A\approx3.69\times10^{-3}$ while for the $\ell=50\,\mu\text m$ case we have $S_A\approx3.49\times10^{-3}$.

The noticeable observation is that the concurrence is now much bigger than when calculated resorting to the Gaussian harmonic oscillator case, see Figures~ \ref{fig:static_concurrence} and \ref{fig:static_concurrence_non_Gaussian}. Moreover, if nature is kind enough, such that the modification of GR at short distances may occur in the micrometer range then we might be able to discern the quantum entanglement of a gravitational system via the QGEM protocol~\cite{Lee_2020}.

Furthermore, we note that when $D, \Delta x \ll \ell$, both $\theta, \phi \rightarrow Gm^2\tau/(2\hbar \ell)$, leading to a vanishing concurrence. In such a case entanglement entropy does not make any sense since the density matrix \eqref{eq:density_matrix_non_gaussian} would be non-invertible. This simple toy model illustrates that non-local gravitational interaction is able to suppress any gravitational-induced entanglement in the deep ultraviolet limit provided all the distance scales, such as the superposition size and the inter separation distance, lie well below the non-local length scale $\ell$.

Of course, realizing a QGEM experiment is a daunting task. Nevertheless, our analysis provides us with a possibility of studying physics beyond the Standard Model in a table-top experiment. There are many experimental challenges, ranging from creating macroscopic superposition for such heavy masses and such a large spatial superposition~\cite{Bose:2017nin,Marshman:2021wyk,Pedernales2020MotionalDD,Zhou:2022frl,Zhou:2022jug,Zhou:2022epb} to reading out the witness~\cite{Tilly:2021qef,Schut:2021svd}, as well as protecting the experiment from jitters and gravity-gradient noise~\cite{Toros:2020dbf,Gunnink:2022ner,Wu:2022rdv}, and various sources of decoherence~\cite{Bose:2017nin,vandeKamp:2020rqh,Rijavec:2020qxd,Schut:2021svd,romero2011large}. Such limitations will not be discussed here since they lie beyond the scope of this investigation.

\section{Conclusions}\label{sec:conclusions}

In this work, we studied the QGEM in the context of perturbative quantum gravity endowed with a non-local interaction with matter. We briefly reviewed the quantization of such a theory following \cite{Gupta}, and we calculated the shift in the gravitational energy when the energy-momentum tensor describes two test particles of equal mass, each one possessing some momentum. We showed that the self-energy contribution is finite for $\ell>0$ and diverges in the local limit $\ell\rightarrow0$. The fact that at the lowest order,
the vacuum energy is always finite
is a property of non-local field theory. In the static limit, the gravitational energy shift is the operator-valued version of the potential derived in \cite{Biswas}.

In order to study the entanglement induced by gravity, we followed \cite{Bose:2022uxe,Marshman:2019sne} and extracted the lowest order matter-matter interaction from the derived operator-valued gravitational energy shift and used this to describe the interaction between two  harmonic traps of equal frequency. We computed the concurrence as well as the von Neumann entropy for the entanglement and compared our results with the $\ell=0$ case. We found that these two quantities decrease when increasing the length scale of non-locality for smaller values of the separation between the two harmonic traps. As expected, for larger values of the aforesaid separation, the concurrence corresponding to different values of $\ell$ coincided. However, the concurrence and von Neumann entropy for the entanglement of the two mesoscopic test masses turned out to be small when compared with the present experimental sensitivity and, therefore, not able to be detected experimentally.

We then turned our attention to examining the setup involving two test masses undergoing parallel spatial splitting based on their spins \cite{Nguyen:2019huk} with the entanglement being induced by the derived gravitational energy shift. For such a scenario, it was also found that increasing the length scale of non-locality resulted in a decrease in the concurrence and von Neumann entropy. In addition, given the magnitude of these measures for the entanglement for this scenario, it is possible that a value for the length scale of non-locality may be found experimentally using this parallel setup, provided nature is kind and the non-local scale is of the order of micrometers, roughly the scale at which the current experiments have probed any modification of GR in a laboratory~\cite{Lee_2020}.

As a final comment, we emphasize that the gravitational theory considered here, which contains an infinite-derivative interaction, is motivated by a special class of infinite-derivative gravity theories containing one analytic and non-zero non-local operator. There are, however, infinite-derivative gravity theories admitting two analytic and non-zero non-local operators. Thus, a possible extension of the present work would be to include a second of such operators and study the resulting entanglement in such a context.

\section*{Acknowledgments}
We would like to thank Antonio L\'opez Maroto for his helpful and insightful discussions. UKBV acknowledges financial support from the National Research Foundation of South Africa, Grant number PMDS22063029733, from the University of Cape Town Postgraduate Funding Office and from the Erasmus+ KA107 International Credit
Mobility Programme. UKBV also acknowledges the hospitality of the IFT/UAM-CSIC Madrid during the latest stages of preparation of the manuscript. AdlCD acknowledges support from Grant PID2021-122938NB-I00 funded by MCIN/AEI/10.13039/501100011033 and by {\it ERDF A way of making Europe} and BG20/00236 action (Ministerio de Universidades, Spain). 
\\ \\ \\ \\

\appendix
\section{Supplementary conditions}\label{sec:supplementary_conditions}
Here we wish to show that, by imposing the specific supplementary conditions suggested in \cite{Gupta,Suraj_N_Gupta_1950}, one is left with two degrees of freedom for physical states associated with the Hamiltonian \eqref{eq:vacuum_hamiltonian} for linearized gravity and that these do not possess negative probabilities. For the case of linearized gravity, the number of degrees of freedom before imposing any supplementary conditions is eleven. For the sake of the ease of notation, in Section \ref{sec:simplified_scenario} below we shall consider a simpler scenario, which arises in the context of Quantum Electrodynamics. For such a case, there are four degrees of freedom, as well as negative norm states, when no supplementary conditions are imposed. Following \cite{Suraj_N_Gupta_1950}, we discuss how the supplementary conditions suggested therein result in there being no physical states with negative norms and that the number of polarizations for such states is reduced to two.

Then, in Section \ref{sec:linearized_gravity_supplementary} we shall apply the arguments of \ref{sec:simplified_scenario} to the case of linearized gravity following \cite{Gupta}. To this end, it is shown following the aforesaid reference that, by imposing similar supplementary conditions, physical states for the case of linearized gravity do not admit negative probabilities and that the number of polarizations reduces from eleven to two. Furthermore, since the energies associated with \eqref{eq:vacuum_hamiltonian} are nonnegative, it is concluded then that the Hamiltonian for linearized gravity is indeed bounded from below.

\subsection{Electromagnetic field}\label{sec:simplified_scenario}
Let us consider the free Maxwell field described by the four-vector potential $A^\mu$. In terms of the creation and annihilation operators, $c^\dagger_\mu\left(\bm k\right)$ and $c_\mu\left(\bm k\right)$ respectively, the four-vector potential may be written as \cite{Suraj_N_Gupta_1950}
\begin{align}\label{eq:A_mu_fourier}
A_\mu=\frac{1}{\left(2\pi\right)^{3/2}}\int\text{d}^3k\frac{1}{\sqrt{2|\bm k|}}\epsilon_\mu\phantom{}^\nu\left[c^\dagger_\nu\left(\bm k\right){\text e}^{-ikx}+c_\nu\left(\bm k\right){\text e}^{ikx}\right]\,,
\end{align}
where $\epsilon^{\mu\nu}$ contains the four polarization vectors as its columns, i.e., for each $\nu\in\{0,1,2,3\}$. It is also noted that $\epsilon^{\alpha\mu}\epsilon_\alpha\phantom{}^\nu=\eta^{\mu\nu}$. We let the $\epsilon^{\mu1}$ and $\epsilon^{\mu2}$ be transverse to the momentum $k^\mu$ while taking $\epsilon^{\mu3}$ to be longitudinal. In what follows, we take $\bm k$ to be aligned along the $z$-axis which gives $\epsilon^{\mu\nu}=\delta^{\mu\nu}$.

Imposing the usual commutation relations on $A^\mu$ and its conjugate momentum yields the following commutation relations for the creation and annihilation operators
\begin{align}
\left[c_i\left(\bm k\right),c_j^\dagger\left(\bm k'\right)\right]&=\delta_{ij}\,\delta^{(3)}\left(\bm k-\bm k'\right)\,,\label{eq:commutation_relation_c_i}\\
\left[c_0\left(\bm k\right),c_0^\dagger\left(\bm k'\right)\right]&=-\,\delta^{(3)}\left(\bm k-\bm k'\right)\,,\label{eq:commutation_relation_c_0}
\end{align}
where ${i,j}\in\left\{1,2,3\right\}$ while the Hamiltonian for the system is \cite{Suraj_N_Gupta_1950}
\begin{align}\label{eq:Hamiltonian_test}
\int\text{d}^3k|\bm k|\left[\sum^3_{i=1}c_i^\dagger\left(\bm k\right)c_i\left(\bm k\right)-c_0^\dagger\left(\bm k\right)c_0\left(\bm k\right)\right]\,.
\end{align}
As a result of the minus sign in the commutation relation \eqref{eq:commutation_relation_c_0}, the operator $-\int\text{d}^3k|\bm k|c^\dagger_0\left(\bm k\right)c_0\left(\bm k\right)$ has nonnegative eigenvalues. As a result, the $c_0\left(\bm k\right)c^\dagger_0\left(\bm k\right)$ term in \eqref{eq:Hamiltonian_test} yields a nonnegative contribution to the energy. Nevertheless, the fact that $c_0\left(\bm k\right)$ satisfies a negative commutation relation may result in negative probabilities. Therefore, it is necessary to impose some supplementary conditions to ensure that such negative probabilities do not occur.

In what follows, we use $\ket{p_1}$, $\ket{q_1}$, $\ket{r_1}$ and $\ket{s_1}$ to denote the first excited $c_1\left(\bm k\right)$, $c_2\left(\bm k\right)$, $c_3\left(\bm k\right)$ and $c_0\left(\bm k\right)$ states respectively. Let us start by considering the general state written as a sum over $n_\mu$ excited $c_\mu\left(\bm k\right)$ states
\begin{align}\label{eq:general_state_test}
\ket{\varphi}:=\sum_{n_1,n_2,n_3,n_0}\sum_{\substack{i_1,\dots,i_{n_1}\\j_1,\dots,j_{n_2}\\k_1,\dots,k_{n_3}\\ \ell_1,\dots,\ell_{n_0}}}N_{i_1,\dots,\ell_{n_0}}\ket{p_{i_1},\dots,s_{\ell_{n_0}}}\,,
\end{align}
where the summation is carried out with the conditions: $i_{m_1+1}>i_{m_1}$, $j_{m_2+1}>j_{m_2}$, $k_{m_3+1}>k_{m_3}$, and $\ell_{m_0+1}>\ell_{m_0}$. At this point, let us impose the following supplementary condition \cite{Suraj_N_Gupta_1950}
\begin{align}\label{eq:supplementary_condition_test}
\partial^\mu A_\mu^{(+)}=0\implies\left[c_3\left(\bm k\right)-c_0\left(\bm k\right)\right]\ket{\varphi}=0\,,
\end{align}
where $A_\mu^{(+)}$ refers to only the ${\text e}^{ikx}$ part in equation \eqref{eq:A_mu_fourier} and the second equality is obtained by using the fact that we have aligned $\bm k$ along the $z$-axis. Applying this condition to the state \eqref{eq:general_state_test} and integrating over $\bm k$ yields
\begin{align}\label{eq:solve_coefficients}
n_3^{1/2}N_{i_1,\dots,\ell_{n_0-1}}+n_0^{1/2}N_{i_1,\dots,k_{n_3-1},\dots,\ell_{n_0}}=0\,.
\end{align}
The positive sign in front of the second term is a result of the negative commutation relation \eqref{eq:commutation_relation_c_0}. It can be noted immediately that the state for which $n_0=n_3=0$ satisfies \eqref{eq:solve_coefficients}. In the case where $n_0=n_3=1$, the following state satisfies \eqref{eq:solve_coefficients}
\begin{align}\label{eq:n_non_zero_state}
\sum_{\substack{i_1,\dots,i_{n_1}\\j_1,\dots,j_{n_2}}}N_{i_1,\dots,j_{n_2}}\left(\ket{p_{i_1},\dots,0,s_1}-\ket{p_{i_1},\dots,r_1,0}\right)\,.
\end{align}
However, the result obtained when acting the Hamiltonian \eqref{eq:Hamiltonian_test} on this state \eqref{eq:n_non_zero_state} is zero. Therefore, the state \eqref{eq:n_non_zero_state} does not result in any observable effect. In fact, this is true for all $n_0,n_3>0$ \cite{Suraj_N_Gupta_1950}. It follows that we may set $n_0=n_3=0$ and, therefore, have no $c_0\left(\bm k\right)$ or $c_3\left(\bm k\right)$ excited states. That is, we have only $c_1\left(\bm k\right)$ and $c_2\left(\bm k\right)$ excitations for physical states associated with the Hamiltonian \eqref{eq:Hamiltonian_test} for the electromagnetic field; leaving us with only the two transverse polarizations. Lastly, we note that since only $c_0\left(\bm k\right)$ satisfies a negative commutation relation and we have $n_0=0$, there will be no negative norm states and, thus, the Hamiltonian \eqref{eq:Hamiltonian_test} is bounded from below.

\subsection{Linearized gravity}\label{sec:linearized_gravity_supplementary}
Let us now return our attention to the system described by the Hamiltonian \eqref{eq:vacuum_hamiltonian}. We define the general state $\ket{\psi}$ as the sum over $n_{\mu\nu}$ $a_{\mu\nu}\left(\bm k\right)$ excitations and $n$ $b\left(\bm k\right)$ excitations. As discussed in Section \ref{sec:quantisation}, the action of the Hamiltonian \eqref{eq:vacuum_hamiltonian} on such a state yields nonnegative energy values. However, in order for the Hamiltonian to be bounded from below, it is necessary to introduce some supplementary conditions to ensure that there are no negative norm states. As suggested in \cite{Gupta}, we impose the supplementary condition
\begin{align}\label{eq:supplementary_condition_1}
\partial^\mu\gamma_{\mu\nu}^{(+)}\ket{\psi}=0\,,
\end{align}
where $\gamma_{\mu\nu}^{(+)}$ refers to only the ${\text e}^{ikx}$ part in \eqref{eq:gamma_mu_nu_Fourier}. By aligning $\bm k$ along the $z$-axis, the supplementary condition \eqref{eq:supplementary_condition_1} yields the following constraints
\begin{align}
\left[a_{00}\left(\bm k\right)-a_{03}\left(\bm k\right)\right]\ket{\psi}=0\,,\label{eq:supplementary_a_00}\\
\left[a_{01}\left(\bm k\right)-a_{13}\left(\bm k\right)\right]\ket{\psi}=0\,,\label{eq:supplementary_a_01}\\
\left[a_{02}\left(\bm k\right)-a_{23}\left(\bm k\right)\right]\ket{\psi}=0\,,\label{eq:supplementary_a_02}\\
\left[a_{03}\left(\bm k\right)-a_{33}\left(\bm k\right)\right]\ket{\psi}=0\,.\label{eq:supplementary_a_03}
\end{align}
Let us start by considering the condition \eqref{eq:supplementary_a_00}. From equation \eqref{eq:commutation_a_mu_nu}, it can be noted that $a_{00}\left(\bm k\right)$ satisfies a positive commutation relation while $a_{03}\left(\bm k\right)$ satisfies a negative one. It follows that this condition is of the same form as \eqref{eq:supplementary_condition_test}. We therefore conclude that any state satisfying \eqref{eq:supplementary_a_00} and involving $a_{00}\left(\bm k\right)$ or $a_{03}\left(\bm k\right)$ excitations will be redundant, i.e., the action of the Hamiltonian \eqref{eq:vacuum_hamiltonian} on such a state is zero. Therefore, we may set $n_{00}=n_{03}=0$; reducing the number of polarizations from eleven to nine. We now turn our attention to the condition \eqref{eq:supplementary_a_01}. From \eqref{eq:commutation_a_mu_nu} it follows that $a_{01}\left(\bm k\right)$ satisfies a negative commutation relation while $a_{13}\left(\bm k\right)$ satisfies a positive one. Therefore, this condition is also of the same form as \eqref{eq:supplementary_condition_test} and we set $n_{01}=n_{13}=0$ since any $a_{01}\left(\bm k\right)$ or $a_{13}\left(\bm k\right)$ excitations would lead to redundant contributions. The condition \eqref{eq:supplementary_a_01} thus reduces the number of polarizations from nine to seven. The same argument may be applied to \eqref{eq:supplementary_a_02} which yields $n_{02}=n_{23}=0$ and the number of polarizations is reduced from seven to five. Applying the same argument to \eqref{eq:supplementary_a_03} results in $n_{03}=n_{33}=0$. However, we already have $n_{03}=0$ as a result of \eqref{eq:supplementary_a_00}. Therefore, the condition \eqref{eq:supplementary_a_03} reduces the number of polarizations from five to four.

The second supplementary condition imposed is \cite{Gupta}
\begin{align}\label{eq:supplementary_condition_2}
\left[\gamma_\mu^{(+)}\phantom{}^\mu-\gamma^{(+)}\right]\ket{\psi}=0\,,
\end{align}
where $\gamma^{(+)}$ refers to only the $\text{\ e}^{ikx}$ part in \eqref{eq:gamma_Fourier}. By defining
\begin{align}
\sqrt2a'_{11}:=a_{11}-a_{22}\,,\ \ \ \ \ \ \ \ \ \sqrt2a'_{22}:=a_{11}+a_{22}\,,
\end{align}
and making use of the fact that there are no $a_{00}\left(\bm k\right)$ or $a_{33}\left(\bm k\right)$ excited states, which follows from the first supplementary condition \eqref{eq:supplementary_condition_1}, the second supplementary condition \eqref{eq:supplementary_condition_2} yields
\begin{align}
\left[\sqrt2a'_{22}\left(\bm k\right)-b\left(\bm k\right)\right]\ket{\psi}=0\,.
\end{align}
The operator $a'_{22}\left(\bm k\right)$ satisfies a positive commutation relation while $b\left(\bm k\right)$ satisfies a negative one. Therefore, it follows from the discussion given in Section \ref{sec:simplified_scenario} that there are no $a'_{22}\left(\bm k\right)$ or $b\left(\bm k\right)$ excited states. The number of polarizations for the physical state $\ket{\psi}$ is now reduced from four to two; described by $a'_{11}\left(\bm k\right)$ and $a_{12}\left(\bm k\right)$ excitations. Moreover since there are no $a_{0i}\left(\bm k\right)$ or $b\left(\bm k\right)$ excitations, which carry negative commutation relations, there are no physical states with negative norms. It therefore follows that the Hamiltonian \eqref{eq:vacuum_hamiltonian} is bounded from below provided that the supplementary conditions \eqref{eq:supplementary_condition_1} and \eqref{eq:supplementary_condition_2} are imposed.
\section{Derivation of equations \eqref{eq:non_static_self_energy} and \eqref{eq:non_static_H_D}}\label{sec:appendix}

In order to evaluate the right-hand side of equation \eqref{eq:hammy_integral_non_static_to_calculate}, we require the integrals
\begin{align}
\int\text{d}^3k\frac{\mathcal{T}^{\dagger}_{00}(\bm k)\mathcal{T}_{00}(\bm k)}{\bm k^2}&=\frac{1}{(2\pi)^3}\int\text{d}^3k\frac{\left( E_A^2+ E_B^2\right){\text e}^{-\ell^2\bm k^2}}{\bm k^2}\nonumber\\
&+\frac{2 E_A E_B}{(2\pi)^3}\int\text{d}^3k\frac{{\text e}^{i\bm k\cdot\left({\bm r}_A-{\bm r}_B\right)-\ell^2\bm k^2}}{\bm k^2}\nonumber\\
=\frac{\left( E_A^2+ E_B^2\right)}{4\pi^{3/2}\ell}&+\frac{ E_A E_B}{2\pi|{\bm r}_A-{\bm r}_B|}\text{erf}\left(\frac{|{\bm r}_A-{\bm r}_B|}{2\ell}\right)\,,\label{eq:T_00_T_00}\\
\int\text{d}^3k\frac{\mathcal{T}^{\dagger}_{03}(\bm k)\mathcal{T}_{03}(\bm k)}{\bm k^2}&=\frac{1}{(2\pi)^3}\int\text{d}^3k\frac{\left( p_A^2+ p_B^2\right){\text e}^{-\ell^2\bm k^2}}{\bm k^2}\nonumber\\
&+\frac{2 p_A p_B}{(2\pi)^3}\int\text{d}^3k\frac{{\text e}^{i\bm k\cdot\left({\bm r}_A-{\bm r}_B\right)-\ell^2\bm k^2}}{\bm k^2}\nonumber\\
=\frac{\left( p_A^2+ p_B^2\right)}{4\pi^{3/2}\ell}&+\frac{ p_A p_B}{2\pi|{\bm r}_A-{\bm r}_B|}\text{erf}\left(\frac{|{\bm r}_A-{\bm r}_B|}{2\ell}\right),\\
\int\text{d}^3k\frac{\mathcal{T}^{\dagger}_{33}(\bm k)\mathcal{T}_{33}(\bm k)}{\bm k^2}
&=\frac{1}{4\pi^{3/2}\ell}\left(\frac{ p_A^4}{ E_A^2}+\frac{ p^4_B}{ E_B^2}\right)\nonumber\\
+\frac{ p^2_A p^2_B}{2\pi  E_A E_B|{\bm r}_A-{\bm r}_B|}&\text{erf}\left(\frac{|{\bm r}_A-{\bm r}_B|}{2\ell}\right),
\end{align}
\begin{align}
&\int\text{d}^3k\frac{\mathcal{T}^{\dagger}_{00}(\bm k)\mathcal{T}_{33}(\bm k)}{\bm k^2}
=\int\text{d}^3k\frac{\mathcal{T}^{\dagger}_{33}(\bm k)\mathcal{T}_{00}(\bm k)}{\bm k^2}\nonumber\\
&=\frac{\left( p_A^2+ p^2_B\right)}{4\pi^{3/2}\ell}\nonumber\\
&+\frac{1}{4\pi|{\bm r}_A-{\bm r}_B|}\left(\frac{ E_A p_B^2}{ E_B}+\frac{ E_B p_A^2}{ E_A}\right)\text{erf}\left(\frac{|{\bm r}_A-{\bm r}_B|}{2\ell}\right).\label{eq:T_33_T_33}
\end{align}
By substituting equations \eqref{eq:T_00_T_00}-\eqref{eq:T_33_T_33} into \eqref{eq:hammy_integral_non_static_to_calculate} one can obtain the gravitational energy shift $\Delta H$. We note that the self-energy \eqref{eq:non_static_self_energy} is made up of the first terms on the right-hand sides of equations \eqref{eq:T_00_T_00}-\eqref{eq:T_33_T_33} while the second terms give the interaction energy \eqref{eq:non_static_H_D}.
\bibliographystyle{apsrev4-1}
\bibliography{IDG_QGEM_bib}

\begin{thebibliography}{93}%
\makeatletter
\providecommand \@ifxundefined [1]{%
 \@ifx{#1\undefined}
}%
\providecommand \@ifnum [1]{%
 \ifnum #1\expandafter \@firstoftwo
 \else \expandafter \@secondoftwo
 \fi
}%
\providecommand \@ifx [1]{%
 \ifx #1\expandafter \@firstoftwo
 \else \expandafter \@secondoftwo
 \fi
}%
\providecommand \natexlab [1]{#1}%
\providecommand \enquote  [1]{``#1''}%
\providecommand \bibnamefont  [1]{#1}%
\providecommand \bibfnamefont [1]{#1}%
\providecommand \citenamefont [1]{#1}%
\providecommand \href@noop [0]{\@secondoftwo}%
\providecommand \href [0]{\begingroup \@sanitize@url \@href}%
\providecommand \@href[1]{\@@startlink{#1}\@@href}%
\providecommand \@@href[1]{\endgroup#1\@@endlink}%
\providecommand \@sanitize@url [0]{\catcode `\\12\catcode `\$12\catcode
  `\&12\catcode `\#12\catcode `\^12\catcode `\_12\catcode `\%12\relax}%
\providecommand \@@startlink[1]{}%
\providecommand \@@endlink[0]{}%
\providecommand \url  [0]{\begingroup\@sanitize@url \@url }%
\providecommand \@url [1]{\endgroup\@href {#1}{\urlprefix }}%
\providecommand \urlprefix  [0]{URL }%
\providecommand \Eprint [0]{\href }%
\providecommand \doibase [0]{http://dx.doi.org/}%
\providecommand \selectlanguage [0]{\@gobble}%
\providecommand \bibinfo  [0]{\@secondoftwo}%
\providecommand \bibfield  [0]{\@secondoftwo}%
\providecommand \translation [1]{[#1]}%
\providecommand \BibitemOpen [0]{}%
\providecommand \bibitemStop [0]{}%
\providecommand \bibitemNoStop [0]{.\EOS\space}%
\providecommand \EOS [0]{\spacefactor3000\relax}%
\providecommand \BibitemShut  [1]{\csname bibitem#1\endcsname}%
\let\auto@bib@innerbib\@empty
\bibitem [{\citenamefont {Will}(2014)}]{Will:2014kxa}%
  \BibitemOpen
  \bibfield  {author} {\bibinfo {author} {\bibfnamefont {C.~M.}\ \bibnamefont
  {Will}},\ }\href {\doibase 10.12942/lrr-2014-4} {\bibfield  {journal}
  {\bibinfo  {journal} {Living Rev. Rel.}\ }\textbf {\bibinfo {volume} {17}},\
  \bibinfo {pages} {4} (\bibinfo {year} {2014})},\ \Eprint
  {http://arxiv.org/abs/1403.7377} {arXiv:1403.7377 [gr-qc]} \BibitemShut
  {NoStop}%
\bibitem [{\citenamefont {Abbott}\ \emph {et~al.}(2016)\citenamefont {Abbott}
  \emph {et~al.}}]{LIGOScientific:2016aoc}%
  \BibitemOpen
  \bibfield  {author} {\bibinfo {author} {\bibfnamefont {B.~P.}\ \bibnamefont
  {Abbott}} \emph {et~al.} (\bibinfo {collaboration} {LIGO Scientific,
  Virgo}),\ }\href {\doibase 10.1103/PhysRevLett.116.061102} {\bibfield
  {journal} {\bibinfo  {journal} {Phys. Rev. Lett.}\ }\textbf {\bibinfo
  {volume} {116}},\ \bibinfo {pages} {061102} (\bibinfo {year} {2016})},\
  \Eprint {http://arxiv.org/abs/1602.03837} {arXiv:1602.03837 [gr-qc]}
  \BibitemShut {NoStop}%
\bibitem [{\citenamefont {Hawking}\ and\ \citenamefont {Ellis}(1975)}]{ellis}%
  \BibitemOpen
  \bibfield  {author} {\bibinfo {author} {\bibfnamefont {S.}~\bibnamefont
  {Hawking}}\ and\ \bibinfo {author} {\bibfnamefont {G.}~\bibnamefont
  {Ellis}},\ }\href {https://books.google.co.za/books?id=9AUhAwAAQBAJ} {\emph
  {\bibinfo {title} {The Large Scale Structure of Space-Time}}},\ Cambridge
  Monographs on Mathematical Physics\ (\bibinfo  {publisher} {Cambridge
  University Press},\ \bibinfo {address} {Cambridge U.K.},\ \bibinfo {year}
  {1975})\BibitemShut {NoStop}%
\bibitem [{\citenamefont {Kiefer}(2007)}]{Kiefer:2007ria}%
  \BibitemOpen
  \bibfield  {author} {\bibinfo {author} {\bibfnamefont {C.}~\bibnamefont
  {Kiefer}},\ }\href@noop {} {\emph {\bibinfo {title} {{Quantum Gravity}}}},\
  \bibinfo {edition} {2nd}\ ed.\ (\bibinfo  {publisher} {Oxford University
  Press},\ \bibinfo {address} {New York},\ \bibinfo {year} {2007})\BibitemShut
  {NoStop}%
\bibitem [{\citenamefont {Bose}\ \emph {et~al.}(2017)\citenamefont {Bose},
  \citenamefont {Mazumdar}, \citenamefont {Morley}, \citenamefont {Ulbricht},
  \citenamefont {Toro\v{s}}, \citenamefont {Paternostro}, \citenamefont
  {Geraci}, \citenamefont {Barker}, \citenamefont {Kim},\ and\ \citenamefont
  {Milburn}}]{Bose:2017nin}%
  \BibitemOpen
  \bibfield  {author} {\bibinfo {author} {\bibfnamefont {S.}~\bibnamefont
  {Bose}}, \bibinfo {author} {\bibfnamefont {A.}~\bibnamefont {Mazumdar}},
  \bibinfo {author} {\bibfnamefont {G.~W.}\ \bibnamefont {Morley}}, \bibinfo
  {author} {\bibfnamefont {H.}~\bibnamefont {Ulbricht}}, \bibinfo {author}
  {\bibfnamefont {M.}~\bibnamefont {Toro\v{s}}}, \bibinfo {author}
  {\bibfnamefont {M.}~\bibnamefont {Paternostro}}, \bibinfo {author}
  {\bibfnamefont {A.}~\bibnamefont {Geraci}}, \bibinfo {author} {\bibfnamefont
  {P.}~\bibnamefont {Barker}}, \bibinfo {author} {\bibfnamefont {M.~S.}\
  \bibnamefont {Kim}}, \ and\ \bibinfo {author} {\bibfnamefont
  {G.}~\bibnamefont {Milburn}},\ }\href {\doibase
  10.1103/PhysRevLett.119.240401} {\bibfield  {journal} {\bibinfo  {journal}
  {Phys. Rev. Lett.}\ }\textbf {\bibinfo {volume} {119}},\ \bibinfo {pages}
  {240401} (\bibinfo {year} {2017})},\ \Eprint
  {http://arxiv.org/abs/1707.06050} {arXiv:1707.06050 [quant-ph]} \BibitemShut
  {NoStop}%
\bibitem [{ICT(2016)}]{ICTS}%
  \BibitemOpen
  \href@noop {} {}\bibinfo {howpublished}
  {\url{https://www.youtube.com/watch?v=0Fv-0k13s_k}} (\bibinfo {year}
  {2016}),\ \bibinfo {note} {accessed 1/11/22}\BibitemShut {NoStop}%
\bibitem [{\citenamefont {Marletto}\ and\ \citenamefont
  {Vedral}(2017)}]{marletto2017gravitationally}%
  \BibitemOpen
  \bibfield  {author} {\bibinfo {author} {\bibfnamefont {C.}~\bibnamefont
  {Marletto}}\ and\ \bibinfo {author} {\bibfnamefont {V.}~\bibnamefont
  {Vedral}},\ }\href@noop {} {\bibfield  {journal} {\bibinfo  {journal}
  {Physical Review Letters}\ }\textbf {\bibinfo {volume} {119}},\ \bibinfo
  {pages} {240402} (\bibinfo {year} {2017})}\BibitemShut {NoStop}%
\bibitem [{\citenamefont {Biswas}\ \emph {et~al.}(2022)\citenamefont {Biswas},
  \citenamefont {Bose}, \citenamefont {Mazumdar},\ and\ \citenamefont
  {Toro\v{s}}}]{Biswas:2022qto}%
  \BibitemOpen
  \bibfield  {author} {\bibinfo {author} {\bibfnamefont {D.}~\bibnamefont
  {Biswas}}, \bibinfo {author} {\bibfnamefont {S.}~\bibnamefont {Bose}},
  \bibinfo {author} {\bibfnamefont {A.}~\bibnamefont {Mazumdar}}, \ and\
  \bibinfo {author} {\bibfnamefont {M.}~\bibnamefont {Toro\v{s}}},\ }\href@noop
  {} {\  (\bibinfo {year} {2022})},\ \Eprint {http://arxiv.org/abs/2209.09273}
  {arXiv:2209.09273 [gr-qc]} \BibitemShut {NoStop}%
\bibitem [{\citenamefont {Bennett}\ \emph {et~al.}(1996)\citenamefont
  {Bennett}, \citenamefont {DiVincenzo}, \citenamefont {Smolin},\ and\
  \citenamefont {Wootters}}]{Bennett_1996}%
  \BibitemOpen
  \bibfield  {author} {\bibinfo {author} {\bibfnamefont {C.~H.}\ \bibnamefont
  {Bennett}}, \bibinfo {author} {\bibfnamefont {D.~P.}\ \bibnamefont
  {DiVincenzo}}, \bibinfo {author} {\bibfnamefont {J.~A.}\ \bibnamefont
  {Smolin}}, \ and\ \bibinfo {author} {\bibfnamefont {W.~K.}\ \bibnamefont
  {Wootters}},\ }\href {\doibase 10.1103/physreva.54.3824} {\bibfield
  {journal} {\bibinfo  {journal} {Physical Review A}\ }\textbf {\bibinfo
  {volume} {54}},\ \bibinfo {pages} {3824} (\bibinfo {year}
  {1996})}\BibitemShut {NoStop}%
\bibitem [{\citenamefont {Hill}\ and\ \citenamefont
  {Wootters}(1997)}]{wooters1}%
  \BibitemOpen
  \bibfield  {author} {\bibinfo {author} {\bibfnamefont {S.}~\bibnamefont
  {Hill}}\ and\ \bibinfo {author} {\bibfnamefont {W.~K.}\ \bibnamefont
  {Wootters}},\ }\href {\doibase 10.1103/PhysRevLett.78.5022} {\bibfield
  {journal} {\bibinfo  {journal} {Phys. Rev. Lett.}\ }\textbf {\bibinfo
  {volume} {78}},\ \bibinfo {pages} {5022} (\bibinfo {year} {1997})},\ \Eprint
  {http://arxiv.org/abs/quant-ph/9703041} {arXiv:quant-ph/9703041} \BibitemShut
  {NoStop}%
\bibitem [{\citenamefont {Marshman}\ \emph {et~al.}(2020)\citenamefont
  {Marshman}, \citenamefont {Mazumdar},\ and\ \citenamefont
  {Bose}}]{Marshman:2019sne}%
  \BibitemOpen
  \bibfield  {author} {\bibinfo {author} {\bibfnamefont {R.~J.}\ \bibnamefont
  {Marshman}}, \bibinfo {author} {\bibfnamefont {A.}~\bibnamefont {Mazumdar}},
  \ and\ \bibinfo {author} {\bibfnamefont {S.}~\bibnamefont {Bose}},\ }\href
  {\doibase 10.1103/PhysRevA.101.052110} {\bibfield  {journal} {\bibinfo
  {journal} {Phys. Rev. A}\ }\textbf {\bibinfo {volume} {101}},\ \bibinfo
  {pages} {052110} (\bibinfo {year} {2020})},\ \Eprint
  {http://arxiv.org/abs/1907.01568} {arXiv:1907.01568 [quant-ph]} \BibitemShut
  {NoStop}%
\bibitem [{\citenamefont {Bose}\ \emph {et~al.}(2022)\citenamefont {Bose},
  \citenamefont {Mazumdar}, \citenamefont {Schut},\ and\ \citenamefont
  {Toro\v{s}}}]{Bose:2022uxe}%
  \BibitemOpen
  \bibfield  {author} {\bibinfo {author} {\bibfnamefont {S.}~\bibnamefont
  {Bose}}, \bibinfo {author} {\bibfnamefont {A.}~\bibnamefont {Mazumdar}},
  \bibinfo {author} {\bibfnamefont {M.}~\bibnamefont {Schut}}, \ and\ \bibinfo
  {author} {\bibfnamefont {M.}~\bibnamefont {Toro\v{s}}},\ }\href {\doibase
  10.1103/PhysRevD.105.106028} {\bibfield  {journal} {\bibinfo  {journal}
  {Phys. Rev. D}\ }\textbf {\bibinfo {volume} {105}},\ \bibinfo {pages}
  {106028} (\bibinfo {year} {2022})},\ \Eprint
  {http://arxiv.org/abs/2201.03583} {arXiv:2201.03583 [gr-qc]} \BibitemShut
  {NoStop}%
\bibitem [{\citenamefont {Carney}\ \emph {et~al.}(2019)\citenamefont {Carney},
  \citenamefont {Stamp},\ and\ \citenamefont {Taylor}}]{Carney_2019}%
  \BibitemOpen
  \bibfield  {author} {\bibinfo {author} {\bibfnamefont {D.}~\bibnamefont
  {Carney}}, \bibinfo {author} {\bibfnamefont {P.~C.~E.}\ \bibnamefont
  {Stamp}}, \ and\ \bibinfo {author} {\bibfnamefont {J.~M.}\ \bibnamefont
  {Taylor}},\ }\href {\doibase 10.1088/1361-6382/aaf9ca} {\bibfield  {journal}
  {\bibinfo  {journal} {Classical and Quantum Gravity}\ }\textbf {\bibinfo
  {volume} {36}},\ \bibinfo {pages} {034001} (\bibinfo {year}
  {2019})}\BibitemShut {NoStop}%
\bibitem [{\citenamefont {Belenchia}\ \emph {et~al.}(2018)\citenamefont
  {Belenchia}, \citenamefont {Wald}, \citenamefont {Giacomini}, \citenamefont
  {Castro-Ruiz}, \citenamefont {Brukner},\ and\ \citenamefont
  {Aspelmeyer}}]{Belenchia:2018szb}%
  \BibitemOpen
  \bibfield  {author} {\bibinfo {author} {\bibfnamefont {A.}~\bibnamefont
  {Belenchia}}, \bibinfo {author} {\bibfnamefont {R.~M.}\ \bibnamefont {Wald}},
  \bibinfo {author} {\bibfnamefont {F.}~\bibnamefont {Giacomini}}, \bibinfo
  {author} {\bibfnamefont {E.}~\bibnamefont {Castro-Ruiz}}, \bibinfo {author}
  {\bibfnamefont {C.}~\bibnamefont {Brukner}}, \ and\ \bibinfo {author}
  {\bibfnamefont {M.}~\bibnamefont {Aspelmeyer}},\ }\href {\doibase
  10.1103/PhysRevD.98.126009} {\bibfield  {journal} {\bibinfo  {journal} {Phys.
  Rev. D}\ }\textbf {\bibinfo {volume} {98}},\ \bibinfo {pages} {126009}
  (\bibinfo {year} {2018})},\ \Eprint {http://arxiv.org/abs/1807.07015}
  {arXiv:1807.07015 [quant-ph]} \BibitemShut {NoStop}%
\bibitem [{\citenamefont {Danielson}\ \emph {et~al.}(2022)\citenamefont
  {Danielson}, \citenamefont {Satishchandran},\ and\ \citenamefont
  {Wald}}]{Danielson:2021egj}%
  \BibitemOpen
  \bibfield  {author} {\bibinfo {author} {\bibfnamefont {D.~L.}\ \bibnamefont
  {Danielson}}, \bibinfo {author} {\bibfnamefont {G.}~\bibnamefont
  {Satishchandran}}, \ and\ \bibinfo {author} {\bibfnamefont {R.~M.}\
  \bibnamefont {Wald}},\ }\href {\doibase 10.1103/PhysRevD.105.086001}
  {\bibfield  {journal} {\bibinfo  {journal} {Phys. Rev. D}\ }\textbf {\bibinfo
  {volume} {105}},\ \bibinfo {pages} {086001} (\bibinfo {year} {2022})},\
  \Eprint {http://arxiv.org/abs/2112.10798} {arXiv:2112.10798 [quant-ph]}
  \BibitemShut {NoStop}%
\bibitem [{\citenamefont {Christodoulou}\ \emph {et~al.}(2023)\citenamefont
  {Christodoulou}, \citenamefont {Di~Biagio}, \citenamefont {Aspelmeyer},
  \citenamefont {Brukner}, \citenamefont {Rovelli},\ and\ \citenamefont
  {Howl}}]{Christodoulou:2022mkf}%
  \BibitemOpen
  \bibfield  {author} {\bibinfo {author} {\bibfnamefont {M.}~\bibnamefont
  {Christodoulou}}, \bibinfo {author} {\bibfnamefont {A.}~\bibnamefont
  {Di~Biagio}}, \bibinfo {author} {\bibfnamefont {M.}~\bibnamefont
  {Aspelmeyer}}, \bibinfo {author} {\bibfnamefont {v.}~\bibnamefont {Brukner}},
  \bibinfo {author} {\bibfnamefont {C.}~\bibnamefont {Rovelli}}, \ and\
  \bibinfo {author} {\bibfnamefont {R.}~\bibnamefont {Howl}},\ }\href {\doibase
  10.1103/PhysRevLett.130.100202} {\bibfield  {journal} {\bibinfo  {journal}
  {Phys. Rev. Lett.}\ }\textbf {\bibinfo {volume} {130}},\ \bibinfo {pages}
  {100202} (\bibinfo {year} {2023})},\ \Eprint
  {http://arxiv.org/abs/2202.03368} {arXiv:2202.03368 [quant-ph]} \BibitemShut
  {NoStop}%
\bibitem [{\citenamefont {Hensen}\ \emph {et~al.}(2015)\citenamefont {Hensen}
  \emph {et~al.}}]{Hensen:2015ccp}%
  \BibitemOpen
  \bibfield  {author} {\bibinfo {author} {\bibfnamefont {B.}~\bibnamefont
  {Hensen}} \emph {et~al.},\ }\href {\doibase 10.1038/nature15759} {\bibfield
  {journal} {\bibinfo  {journal} {Nature}\ }\textbf {\bibinfo {volume} {526}},\
  \bibinfo {pages} {682} (\bibinfo {year} {2015})},\ \Eprint
  {http://arxiv.org/abs/1508.05949} {arXiv:1508.05949 [quant-ph]} \BibitemShut
  {NoStop}%
\bibitem [{\citenamefont {Peres}(1992)}]{PhysRevA.46.4413}%
  \BibitemOpen
  \bibfield  {author} {\bibinfo {author} {\bibfnamefont {A.}~\bibnamefont
  {Peres}},\ }\href {\doibase 10.1103/PhysRevA.46.4413} {\bibfield  {journal}
  {\bibinfo  {journal} {Phys. Rev. A}\ }\textbf {\bibinfo {volume} {46}},\
  \bibinfo {pages} {4413} (\bibinfo {year} {1992})}\BibitemShut {NoStop}%
\bibitem [{\citenamefont {Gisin}\ and\ \citenamefont
  {Peres}(1992)}]{GISIN199215}%
  \BibitemOpen
  \bibfield  {author} {\bibinfo {author} {\bibfnamefont {N.}~\bibnamefont
  {Gisin}}\ and\ \bibinfo {author} {\bibfnamefont {A.}~\bibnamefont {Peres}},\
  }\href {\doibase https://doi.org/10.1016/0375-9601(92)90949-M} {\bibfield
  {journal} {\bibinfo  {journal} {Physics Letters A}\ }\textbf {\bibinfo
  {volume} {162}},\ \bibinfo {pages} {15} (\bibinfo {year} {1992})}\BibitemShut
  {NoStop}%
\bibitem [{\citenamefont {Tomboulis}(2015)}]{Tomboulis:2015esa}%
  \BibitemOpen
  \bibfield  {author} {\bibinfo {author} {\bibfnamefont {E.~T.}\ \bibnamefont
  {Tomboulis}},\ }\href {\doibase 10.1142/S0217732315400052} {\bibfield
  {journal} {\bibinfo  {journal} {Mod. Phys. Lett. A}\ }\textbf {\bibinfo
  {volume} {30}},\ \bibinfo {pages} {1540005} (\bibinfo {year}
  {2015})}\BibitemShut {NoStop}%
\bibitem [{\citenamefont {Buoninfante}\ \emph {et~al.}(2019)\citenamefont
  {Buoninfante}, \citenamefont {Lambiase},\ and\ \citenamefont
  {Mazumdar}}]{Buoninfante:2018mre}%
  \BibitemOpen
  \bibfield  {author} {\bibinfo {author} {\bibfnamefont {L.}~\bibnamefont
  {Buoninfante}}, \bibinfo {author} {\bibfnamefont {G.}~\bibnamefont
  {Lambiase}}, \ and\ \bibinfo {author} {\bibfnamefont {A.}~\bibnamefont
  {Mazumdar}},\ }\href {\doibase 10.1016/j.nuclphysb.2019.114646} {\bibfield
  {journal} {\bibinfo  {journal} {Nucl. Phys. B}\ }\textbf {\bibinfo {volume}
  {944}},\ \bibinfo {pages} {114646} (\bibinfo {year} {2019})},\ \Eprint
  {http://arxiv.org/abs/1805.03559} {arXiv:1805.03559 [hep-th]} \BibitemShut
  {NoStop}%
\bibitem [{\citenamefont {Edholm}\ \emph {et~al.}(2016)\citenamefont {Edholm},
  \citenamefont {Koshelev},\ and\ \citenamefont {Mazumdar}}]{Edholm:2016hbt}%
  \BibitemOpen
  \bibfield  {author} {\bibinfo {author} {\bibfnamefont {J.}~\bibnamefont
  {Edholm}}, \bibinfo {author} {\bibfnamefont {A.~S.}\ \bibnamefont
  {Koshelev}}, \ and\ \bibinfo {author} {\bibfnamefont {A.}~\bibnamefont
  {Mazumdar}},\ }\href {\doibase 10.1103/PhysRevD.94.104033} {\bibfield
  {journal} {\bibinfo  {journal} {Phys. Rev. D}\ }\textbf {\bibinfo {volume}
  {94}},\ \bibinfo {pages} {104033} (\bibinfo {year} {2016})},\ \Eprint
  {http://arxiv.org/abs/1604.01989} {arXiv:1604.01989 [gr-qc]} \BibitemShut
  {NoStop}%
\bibitem [{\citenamefont {Krasnikov}(1987)}]{Krasnikov:1987yj}%
  \BibitemOpen
  \bibfield  {author} {\bibinfo {author} {\bibfnamefont {N.~V.}\ \bibnamefont
  {Krasnikov}},\ }\href {\doibase 10.1007/BF01017588} {\bibfield  {journal}
  {\bibinfo  {journal} {Theor. Math. Phys.}\ }\textbf {\bibinfo {volume}
  {73}},\ \bibinfo {pages} {1184} (\bibinfo {year} {1987})}\BibitemShut
  {NoStop}%
\bibitem [{\citenamefont {Tomboulis}(1997)}]{Tomboulis:1997gg}%
  \BibitemOpen
  \bibfield  {author} {\bibinfo {author} {\bibfnamefont {E.~T.}\ \bibnamefont
  {Tomboulis}},\ }\href@noop {} {\  (\bibinfo {year} {1997})},\ \Eprint
  {http://arxiv.org/abs/hep-th/9702146} {arXiv:hep-th/9702146} \BibitemShut
  {NoStop}%
\bibitem [{\citenamefont {Siegel}(2003)}]{Siegel:2003vt}%
  \BibitemOpen
  \bibfield  {author} {\bibinfo {author} {\bibfnamefont {W.}~\bibnamefont
  {Siegel}},\ }\href@noop {} {\  (\bibinfo {year} {2003})},\ \Eprint
  {http://arxiv.org/abs/hep-th/0309093} {arXiv:hep-th/0309093} \BibitemShut
  {NoStop}%
\bibitem [{\citenamefont {Biswas}\ \emph {et~al.}(2006)\citenamefont {Biswas},
  \citenamefont {Mazumdar},\ and\ \citenamefont {Siegel}}]{Biswas:2005qr}%
  \BibitemOpen
  \bibfield  {author} {\bibinfo {author} {\bibfnamefont {T.}~\bibnamefont
  {Biswas}}, \bibinfo {author} {\bibfnamefont {A.}~\bibnamefont {Mazumdar}}, \
  and\ \bibinfo {author} {\bibfnamefont {W.}~\bibnamefont {Siegel}},\ }\href
  {\doibase 10.1088/1475-7516/2006/03/009} {\bibfield  {journal} {\bibinfo
  {journal} {JCAP}\ }\textbf {\bibinfo {volume} {03}},\ \bibinfo {pages} {009}
  (\bibinfo {year} {2006})},\ \Eprint {http://arxiv.org/abs/hep-th/0508194}
  {arXiv:hep-th/0508194} \BibitemShut {NoStop}%
\bibitem [{\citenamefont {Biswas}\ \emph
  {et~al.}(2012{\natexlab{a}})\citenamefont {Biswas}, \citenamefont {Gerwick},
  \citenamefont {Koivisto},\ and\ \citenamefont {Mazumdar}}]{Biswas}%
  \BibitemOpen
  \bibfield  {author} {\bibinfo {author} {\bibfnamefont {T.}~\bibnamefont
  {Biswas}}, \bibinfo {author} {\bibfnamefont {E.}~\bibnamefont {Gerwick}},
  \bibinfo {author} {\bibfnamefont {T.}~\bibnamefont {Koivisto}}, \ and\
  \bibinfo {author} {\bibfnamefont {A.}~\bibnamefont {Mazumdar}},\ }\href
  {\doibase 10.1103/PhysRevLett.108.031101} {\bibfield  {journal} {\bibinfo
  {journal} {Phys. Rev. Lett.}\ }\textbf {\bibinfo {volume} {108}},\ \bibinfo
  {pages} {031101} (\bibinfo {year} {2012}{\natexlab{a}})},\ \Eprint
  {http://arxiv.org/abs/1110.5249} {arXiv:1110.5249 [gr-qc]} \BibitemShut
  {NoStop}%
\bibitem [{\citenamefont {Modesto}(2012)}]{Modesto:2011kw}%
  \BibitemOpen
  \bibfield  {author} {\bibinfo {author} {\bibfnamefont {L.}~\bibnamefont
  {Modesto}},\ }\href {\doibase 10.1103/PhysRevD.86.044005} {\bibfield
  {journal} {\bibinfo  {journal} {Phys. Rev. D}\ }\textbf {\bibinfo {volume}
  {86}},\ \bibinfo {pages} {044005} (\bibinfo {year} {2012})},\ \Eprint
  {http://arxiv.org/abs/1107.2403} {arXiv:1107.2403 [hep-th]} \BibitemShut
  {NoStop}%
\bibitem [{\citenamefont {Deser}\ and\ \citenamefont
  {Woodard}(2007)}]{Deser:2007jk}%
  \BibitemOpen
  \bibfield  {author} {\bibinfo {author} {\bibfnamefont {S.}~\bibnamefont
  {Deser}}\ and\ \bibinfo {author} {\bibfnamefont {R.~P.}\ \bibnamefont
  {Woodard}},\ }\href {\doibase 10.1103/PhysRevLett.99.111301} {\bibfield
  {journal} {\bibinfo  {journal} {Phys. Rev. Lett.}\ }\textbf {\bibinfo
  {volume} {99}},\ \bibinfo {pages} {111301} (\bibinfo {year} {2007})},\
  \Eprint {http://arxiv.org/abs/0706.2151} {arXiv:0706.2151 [astro-ph]}
  \BibitemShut {NoStop}%
\bibitem [{\citenamefont {Woodard}(2014)}]{Woodard:2014iga}%
  \BibitemOpen
  \bibfield  {author} {\bibinfo {author} {\bibfnamefont {R.~P.}\ \bibnamefont
  {Woodard}},\ }\href {\doibase 10.1007/s10701-014-9780-6} {\bibfield
  {journal} {\bibinfo  {journal} {Found. Phys.}\ }\textbf {\bibinfo {volume}
  {44}},\ \bibinfo {pages} {213} (\bibinfo {year} {2014})},\ \Eprint
  {http://arxiv.org/abs/1401.0254} {arXiv:1401.0254 [astro-ph.CO]} \BibitemShut
  {NoStop}%
\bibitem [{\citenamefont {de~Lacroix}\ \emph {et~al.}(2017)\citenamefont
  {de~Lacroix}, \citenamefont {Erbin}, \citenamefont {Kashyap}, \citenamefont
  {Sen},\ and\ \citenamefont {Verma}}]{deLacroix:2017lif}%
  \BibitemOpen
  \bibfield  {author} {\bibinfo {author} {\bibfnamefont {C.}~\bibnamefont
  {de~Lacroix}}, \bibinfo {author} {\bibfnamefont {H.}~\bibnamefont {Erbin}},
  \bibinfo {author} {\bibfnamefont {S.~P.}\ \bibnamefont {Kashyap}}, \bibinfo
  {author} {\bibfnamefont {A.}~\bibnamefont {Sen}}, \ and\ \bibinfo {author}
  {\bibfnamefont {M.}~\bibnamefont {Verma}},\ }\href {\doibase
  10.1142/S0217751X17300216} {\bibfield  {journal} {\bibinfo  {journal} {Int.
  J. Mod. Phys. A}\ }\textbf {\bibinfo {volume} {32}},\ \bibinfo {pages}
  {1730021} (\bibinfo {year} {2017})},\ \Eprint
  {http://arxiv.org/abs/1703.06410} {arXiv:1703.06410 [hep-th]} \BibitemShut
  {NoStop}%
\bibitem [{\citenamefont {Freund}(2017)}]{Freund:2017aqf}%
  \BibitemOpen
  \bibfield  {author} {\bibinfo {author} {\bibfnamefont {P.~G.~O.}\
  \bibnamefont {Freund}}\ }(\bibinfo {year} {2017})\ \Eprint
  {http://arxiv.org/abs/1711.00523} {arXiv:1711.00523 [hep-th]} \BibitemShut
  {NoStop}%
\bibitem [{\citenamefont {Freund}\ and\ \citenamefont
  {Witten}(1987)}]{FREUND1987191}%
  \BibitemOpen
  \bibfield  {author} {\bibinfo {author} {\bibfnamefont {P.~G.}\ \bibnamefont
  {Freund}}\ and\ \bibinfo {author} {\bibfnamefont {E.}~\bibnamefont
  {Witten}},\ }\href {\doibase https://doi.org/10.1016/0370-2693(87)91357-8}
  {\bibfield  {journal} {\bibinfo  {journal} {Physics Letters B}\ }\textbf
  {\bibinfo {volume} {199}},\ \bibinfo {pages} {191} (\bibinfo {year}
  {1987})}\BibitemShut {NoStop}%
\bibitem [{\citenamefont {Abel}\ \emph {et~al.}(2020)\citenamefont {Abel},
  \citenamefont {Buoninfante},\ and\ \citenamefont {Mazumdar}}]{Abel:2019zou}%
  \BibitemOpen
  \bibfield  {author} {\bibinfo {author} {\bibfnamefont {S.}~\bibnamefont
  {Abel}}, \bibinfo {author} {\bibfnamefont {L.}~\bibnamefont {Buoninfante}}, \
  and\ \bibinfo {author} {\bibfnamefont {A.}~\bibnamefont {Mazumdar}},\ }\href
  {\doibase 10.1007/JHEP01(2020)003} {\bibfield  {journal} {\bibinfo  {journal}
  {JHEP}\ }\textbf {\bibinfo {volume} {01}},\ \bibinfo {pages} {003} (\bibinfo
  {year} {2020})},\ \Eprint {http://arxiv.org/abs/1911.06697} {arXiv:1911.06697
  [hep-th]} \BibitemShut {NoStop}%
\bibitem [{\citenamefont {Biswas}\ \emph {et~al.}(2010)\citenamefont {Biswas},
  \citenamefont {Koivisto},\ and\ \citenamefont {Mazumdar}}]{Biswas:2010zk}%
  \BibitemOpen
  \bibfield  {author} {\bibinfo {author} {\bibfnamefont {T.}~\bibnamefont
  {Biswas}}, \bibinfo {author} {\bibfnamefont {T.}~\bibnamefont {Koivisto}}, \
  and\ \bibinfo {author} {\bibfnamefont {A.}~\bibnamefont {Mazumdar}},\ }\href
  {\doibase 10.1088/1475-7516/2010/11/008} {\bibfield  {journal} {\bibinfo
  {journal} {JCAP}\ }\textbf {\bibinfo {volume} {11}},\ \bibinfo {pages} {008}
  (\bibinfo {year} {2010})},\ \Eprint {http://arxiv.org/abs/1005.0590}
  {arXiv:1005.0590 [hep-th]} \BibitemShut {NoStop}%
\bibitem [{\citenamefont {Biswas}\ \emph
  {et~al.}(2012{\natexlab{b}})\citenamefont {Biswas}, \citenamefont {Koshelev},
  \citenamefont {Mazumdar},\ and\ \citenamefont {Vernov}}]{Biswas:2012bp}%
  \BibitemOpen
  \bibfield  {author} {\bibinfo {author} {\bibfnamefont {T.}~\bibnamefont
  {Biswas}}, \bibinfo {author} {\bibfnamefont {A.~S.}\ \bibnamefont
  {Koshelev}}, \bibinfo {author} {\bibfnamefont {A.}~\bibnamefont {Mazumdar}},
  \ and\ \bibinfo {author} {\bibfnamefont {S.~Y.}\ \bibnamefont {Vernov}},\
  }\href {\doibase 10.1088/1475-7516/2012/08/024} {\bibfield  {journal}
  {\bibinfo  {journal} {JCAP}\ }\textbf {\bibinfo {volume} {08}},\ \bibinfo
  {pages} {024} (\bibinfo {year} {2012}{\natexlab{b}})},\ \Eprint
  {http://arxiv.org/abs/1206.6374} {arXiv:1206.6374 [astro-ph.CO]} \BibitemShut
  {NoStop}%
\bibitem [{\citenamefont {Kumar}\ \emph {et~al.}(2021)\citenamefont {Kumar},
  \citenamefont {Maheshwari}, \citenamefont {Mazumdar},\ and\ \citenamefont
  {Peng}}]{Kumar:2021mgc}%
  \BibitemOpen
  \bibfield  {author} {\bibinfo {author} {\bibfnamefont {K.~S.}\ \bibnamefont
  {Kumar}}, \bibinfo {author} {\bibfnamefont {S.}~\bibnamefont {Maheshwari}},
  \bibinfo {author} {\bibfnamefont {A.}~\bibnamefont {Mazumdar}}, \ and\
  \bibinfo {author} {\bibfnamefont {J.}~\bibnamefont {Peng}},\ }\href {\doibase
  10.1088/1475-7516/2021/07/025} {\bibfield  {journal} {\bibinfo  {journal}
  {JCAP}\ }\textbf {\bibinfo {volume} {07}},\ \bibinfo {pages} {025} (\bibinfo
  {year} {2021})},\ \Eprint {http://arxiv.org/abs/2103.13980} {arXiv:2103.13980
  [gr-qc]} \BibitemShut {NoStop}%
\bibitem [{\citenamefont {Frolov}\ \emph {et~al.}(2015)\citenamefont {Frolov},
  \citenamefont {Zelnikov},\ and\ \citenamefont
  {de~Paula~Netto}}]{Frolov:2015bia}%
  \BibitemOpen
  \bibfield  {author} {\bibinfo {author} {\bibfnamefont {V.~P.}\ \bibnamefont
  {Frolov}}, \bibinfo {author} {\bibfnamefont {A.}~\bibnamefont {Zelnikov}}, \
  and\ \bibinfo {author} {\bibfnamefont {T.}~\bibnamefont {de~Paula~Netto}},\
  }\href {\doibase 10.1007/JHEP06(2015)107} {\bibfield  {journal} {\bibinfo
  {journal} {JHEP}\ }\textbf {\bibinfo {volume} {06}},\ \bibinfo {pages} {107}
  (\bibinfo {year} {2015})},\ \Eprint {http://arxiv.org/abs/1504.00412}
  {arXiv:1504.00412 [hep-th]} \BibitemShut {NoStop}%
\bibitem [{\citenamefont {Frolov}\ and\ \citenamefont
  {Zelnikov}(2016)}]{Frolov:2015usa}%
  \BibitemOpen
  \bibfield  {author} {\bibinfo {author} {\bibfnamefont {V.~P.}\ \bibnamefont
  {Frolov}}\ and\ \bibinfo {author} {\bibfnamefont {A.}~\bibnamefont
  {Zelnikov}},\ }\href {\doibase 10.1103/PhysRevD.93.064048} {\bibfield
  {journal} {\bibinfo  {journal} {Phys. Rev. D}\ }\textbf {\bibinfo {volume}
  {93}},\ \bibinfo {pages} {064048} (\bibinfo {year} {2016})},\ \Eprint
  {http://arxiv.org/abs/1509.03336} {arXiv:1509.03336 [hep-th]} \BibitemShut
  {NoStop}%
\bibitem [{\citenamefont {Buoninfante}\ \emph
  {et~al.}(2018{\natexlab{a}})\citenamefont {Buoninfante}, \citenamefont
  {Koshelev}, \citenamefont {Lambiase},\ and\ \citenamefont
  {Mazumdar}}]{Buoninfante:2018xiw}%
  \BibitemOpen
  \bibfield  {author} {\bibinfo {author} {\bibfnamefont {L.}~\bibnamefont
  {Buoninfante}}, \bibinfo {author} {\bibfnamefont {A.~S.}\ \bibnamefont
  {Koshelev}}, \bibinfo {author} {\bibfnamefont {G.}~\bibnamefont {Lambiase}},
  \ and\ \bibinfo {author} {\bibfnamefont {A.}~\bibnamefont {Mazumdar}},\
  }\href {\doibase 10.1088/1475-7516/2018/09/034} {\bibfield  {journal}
  {\bibinfo  {journal} {JCAP}\ }\textbf {\bibinfo {volume} {09}},\ \bibinfo
  {pages} {034} (\bibinfo {year} {2018}{\natexlab{a}})},\ \Eprint
  {http://arxiv.org/abs/1802.00399} {arXiv:1802.00399 [gr-qc]} \BibitemShut
  {NoStop}%
\bibitem [{\citenamefont {Boos}\ \emph {et~al.}(2018)\citenamefont {Boos},
  \citenamefont {Frolov},\ and\ \citenamefont {Zelnikov}}]{Boos:2018bxf}%
  \BibitemOpen
  \bibfield  {author} {\bibinfo {author} {\bibfnamefont {J.}~\bibnamefont
  {Boos}}, \bibinfo {author} {\bibfnamefont {V.~P.}\ \bibnamefont {Frolov}}, \
  and\ \bibinfo {author} {\bibfnamefont {A.}~\bibnamefont {Zelnikov}},\ }\href
  {\doibase 10.1103/PhysRevD.97.084021} {\bibfield  {journal} {\bibinfo
  {journal} {Phys. Rev. D}\ }\textbf {\bibinfo {volume} {97}},\ \bibinfo
  {pages} {084021} (\bibinfo {year} {2018})},\ \Eprint
  {http://arxiv.org/abs/1802.09573} {arXiv:1802.09573 [gr-qc]} \BibitemShut
  {NoStop}%
\bibitem [{\citenamefont {Kol\'a\v{r}}\ and\ \citenamefont
  {Boos}(2021)}]{Kolar:2021oba}%
  \BibitemOpen
  \bibfield  {author} {\bibinfo {author} {\bibfnamefont {I.}~\bibnamefont
  {Kol\'a\v{r}}}\ and\ \bibinfo {author} {\bibfnamefont {J.}~\bibnamefont
  {Boos}},\ }\href {\doibase 10.1103/PhysRevD.103.105004} {\bibfield  {journal}
  {\bibinfo  {journal} {Phys. Rev. D}\ }\textbf {\bibinfo {volume} {103}},\
  \bibinfo {pages} {105004} (\bibinfo {year} {2021})},\ \Eprint
  {http://arxiv.org/abs/2102.07843} {arXiv:2102.07843 [hep-th]} \BibitemShut
  {NoStop}%
\bibitem [{\citenamefont {Vinckers}\ \emph {et~al.}(2022)\citenamefont
  {Vinckers}, \citenamefont {de~la Cruz-Dombriz}, \citenamefont {Kol\'a\v{r}},
  \citenamefont {Maldonado~Torralba},\ and\ \citenamefont
  {Mazumdar}}]{Vinckers:2022the}%
  \BibitemOpen
  \bibfield  {author} {\bibinfo {author} {\bibfnamefont {U.~K.~B.}\
  \bibnamefont {Vinckers}}, \bibinfo {author} {\bibfnamefont {A.}~\bibnamefont
  {de~la Cruz-Dombriz}}, \bibinfo {author} {\bibfnamefont {I.}~\bibnamefont
  {Kol\'a\v{r}}}, \bibinfo {author} {\bibfnamefont {F.~J.}\ \bibnamefont
  {Maldonado~Torralba}}, \ and\ \bibinfo {author} {\bibfnamefont
  {A.}~\bibnamefont {Mazumdar}},\ }\href {\doibase 10.1103/PhysRevD.106.064037}
  {\bibfield  {journal} {\bibinfo  {journal} {Phys. Rev. D}\ }\textbf {\bibinfo
  {volume} {106}},\ \bibinfo {pages} {064037} (\bibinfo {year} {2022})},\
  \Eprint {http://arxiv.org/abs/2206.07111} {arXiv:2206.07111 [gr-qc]}
  \BibitemShut {NoStop}%
\bibitem [{\citenamefont {Kol\'a\v{r}}\ and\ \citenamefont
  {Mazumdar}(2020)}]{Kolar:2020bpo}%
  \BibitemOpen
  \bibfield  {author} {\bibinfo {author} {\bibfnamefont {I.}~\bibnamefont
  {Kol\'a\v{r}}}\ and\ \bibinfo {author} {\bibfnamefont {A.}~\bibnamefont
  {Mazumdar}},\ }\href {\doibase 10.1103/PhysRevD.101.124005} {\bibfield
  {journal} {\bibinfo  {journal} {Phys. Rev. D}\ }\textbf {\bibinfo {volume}
  {101}},\ \bibinfo {pages} {124005} (\bibinfo {year} {2020})},\ \Eprint
  {http://arxiv.org/abs/2004.07613} {arXiv:2004.07613 [gr-qc]} \BibitemShut
  {NoStop}%
\bibitem [{\citenamefont {Frolov}()}]{Frolov:2015bta}%
  \BibitemOpen
  \bibfield  {author} {\bibinfo {author} {\bibfnamefont {V.~P.}\ \bibnamefont
  {Frolov}},\ }\href {\doibase 10.1103/PhysRevLett.115.051102} {\bibfield
  {journal} {\bibinfo  {journal} {Phys. Rev. Lett.}\ }\textbf {\bibinfo
  {volume} {115}},\ \bibinfo {pages} {051102}},\ \Eprint
  {http://arxiv.org/abs/1505.00492} {arXiv:1505.00492 [hep-th]} \BibitemShut
  {NoStop}%
\bibitem [{\citenamefont {Koshelev}\ \emph {et~al.}(2018)\citenamefont
  {Koshelev}, \citenamefont {Marto},\ and\ \citenamefont
  {Mazumdar}}]{Koshelev:2018hpt}%
  \BibitemOpen
  \bibfield  {author} {\bibinfo {author} {\bibfnamefont {A.~S.}\ \bibnamefont
  {Koshelev}}, \bibinfo {author} {\bibfnamefont {J.~a.}\ \bibnamefont {Marto}},
  \ and\ \bibinfo {author} {\bibfnamefont {A.}~\bibnamefont {Mazumdar}},\
  }\href {\doibase 10.1103/PhysRevD.98.064023} {\bibfield  {journal} {\bibinfo
  {journal} {Phys. Rev. D}\ }\textbf {\bibinfo {volume} {98}},\ \bibinfo
  {pages} {064023} (\bibinfo {year} {2018})},\ \Eprint
  {http://arxiv.org/abs/1803.00309} {arXiv:1803.00309 [gr-qc]} \BibitemShut
  {NoStop}%
\bibitem [{\citenamefont {Buoninfante}\ and\ \citenamefont
  {Mazumdar}(2019)}]{Buoninfante:2019swn}%
  \BibitemOpen
  \bibfield  {author} {\bibinfo {author} {\bibfnamefont {L.}~\bibnamefont
  {Buoninfante}}\ and\ \bibinfo {author} {\bibfnamefont {A.}~\bibnamefont
  {Mazumdar}},\ }\href {\doibase 10.1103/PhysRevD.100.024031} {\bibfield
  {journal} {\bibinfo  {journal} {Phys. Rev. D}\ }\textbf {\bibinfo {volume}
  {100}},\ \bibinfo {pages} {024031} (\bibinfo {year} {2019})},\ \Eprint
  {http://arxiv.org/abs/1903.01542} {arXiv:1903.01542 [gr-qc]} \BibitemShut
  {NoStop}%
\bibitem [{\citenamefont {Buoninfante}\ \emph
  {et~al.}(2018{\natexlab{b}})\citenamefont {Buoninfante}, \citenamefont
  {Koshelev}, \citenamefont {Lambiase}, \citenamefont {Marto},\ and\
  \citenamefont {Mazumdar}}]{Buoninfante:2018rlq}%
  \BibitemOpen
  \bibfield  {author} {\bibinfo {author} {\bibfnamefont {L.}~\bibnamefont
  {Buoninfante}}, \bibinfo {author} {\bibfnamefont {A.~S.}\ \bibnamefont
  {Koshelev}}, \bibinfo {author} {\bibfnamefont {G.}~\bibnamefont {Lambiase}},
  \bibinfo {author} {\bibfnamefont {J.~a.}\ \bibnamefont {Marto}}, \ and\
  \bibinfo {author} {\bibfnamefont {A.}~\bibnamefont {Mazumdar}},\ }\href
  {\doibase 10.1088/1475-7516/2018/06/014} {\bibfield  {journal} {\bibinfo
  {journal} {JCAP}\ }\textbf {\bibinfo {volume} {06}},\ \bibinfo {pages} {014}
  (\bibinfo {year} {2018}{\natexlab{b}})},\ \Eprint
  {http://arxiv.org/abs/1804.08195} {arXiv:1804.08195 [gr-qc]} \BibitemShut
  {NoStop}%
\bibitem [{\citenamefont {Buoninfante}\ \emph
  {et~al.}(2018{\natexlab{c}})\citenamefont {Buoninfante}, \citenamefont
  {Cornell}, \citenamefont {Harmsen}, \citenamefont {Koshelev}, \citenamefont
  {Lambiase}, \citenamefont {Marto},\ and\ \citenamefont
  {Mazumdar}}]{Buoninfante:2018xif}%
  \BibitemOpen
  \bibfield  {author} {\bibinfo {author} {\bibfnamefont {L.}~\bibnamefont
  {Buoninfante}}, \bibinfo {author} {\bibfnamefont {A.~S.}\ \bibnamefont
  {Cornell}}, \bibinfo {author} {\bibfnamefont {G.}~\bibnamefont {Harmsen}},
  \bibinfo {author} {\bibfnamefont {A.~S.}\ \bibnamefont {Koshelev}}, \bibinfo
  {author} {\bibfnamefont {G.}~\bibnamefont {Lambiase}}, \bibinfo {author}
  {\bibfnamefont {J.~a.}\ \bibnamefont {Marto}}, \ and\ \bibinfo {author}
  {\bibfnamefont {A.}~\bibnamefont {Mazumdar}},\ }\href {\doibase
  10.1103/PhysRevD.98.084041} {\bibfield  {journal} {\bibinfo  {journal} {Phys.
  Rev. D}\ }\textbf {\bibinfo {volume} {98}},\ \bibinfo {pages} {084041}
  (\bibinfo {year} {2018}{\natexlab{c}})},\ \Eprint
  {http://arxiv.org/abs/1807.08896} {arXiv:1807.08896 [gr-qc]} \BibitemShut
  {NoStop}%
\bibitem [{\citenamefont {Kilicarslan}(2019)}]{Kilicarslan:2019njc}%
  \BibitemOpen
  \bibfield  {author} {\bibinfo {author} {\bibfnamefont {E.}~\bibnamefont
  {Kilicarslan}},\ }\href {\doibase 10.1103/PhysRevD.99.124048} {\bibfield
  {journal} {\bibinfo  {journal} {Phys. Rev. D}\ }\textbf {\bibinfo {volume}
  {99}},\ \bibinfo {pages} {124048} (\bibinfo {year} {2019})},\ \Eprint
  {http://arxiv.org/abs/1903.04283} {arXiv:1903.04283 [gr-qc]} \BibitemShut
  {NoStop}%
\bibitem [{\citenamefont {Dengiz}\ \emph {et~al.}(2020)\citenamefont {Dengiz},
  \citenamefont {Kilicarslan}, \citenamefont {Kol\'a\v{r}},\ and\ \citenamefont
  {Mazumdar}}]{Dengiz:2020xbu}%
  \BibitemOpen
  \bibfield  {author} {\bibinfo {author} {\bibfnamefont {S.}~\bibnamefont
  {Dengiz}}, \bibinfo {author} {\bibfnamefont {E.}~\bibnamefont {Kilicarslan}},
  \bibinfo {author} {\bibfnamefont {I.}~\bibnamefont {Kol\'a\v{r}}}, \ and\
  \bibinfo {author} {\bibfnamefont {A.}~\bibnamefont {Mazumdar}},\ }\href
  {\doibase 10.1103/PhysRevD.102.044016} {\bibfield  {journal} {\bibinfo
  {journal} {Phys. Rev. D}\ }\textbf {\bibinfo {volume} {102}},\ \bibinfo
  {pages} {044016} (\bibinfo {year} {2020})},\ \Eprint
  {http://arxiv.org/abs/2006.07650} {arXiv:2006.07650 [gr-qc]} \BibitemShut
  {NoStop}%
\bibitem [{\citenamefont {Kol\'a\v{r}}\ \emph
  {et~al.}(2021{\natexlab{a}})\citenamefont {Kol\'a\v{r}}, \citenamefont
  {M\'alek},\ and\ \citenamefont {Mazumdar}}]{Kolar:2021rfl}%
  \BibitemOpen
  \bibfield  {author} {\bibinfo {author} {\bibfnamefont {I.}~\bibnamefont
  {Kol\'a\v{r}}}, \bibinfo {author} {\bibfnamefont {T.}~\bibnamefont
  {M\'alek}}, \ and\ \bibinfo {author} {\bibfnamefont {A.}~\bibnamefont
  {Mazumdar}},\ }\href {\doibase 10.1103/PhysRevD.103.124067} {\bibfield
  {journal} {\bibinfo  {journal} {Phys. Rev. D}\ }\textbf {\bibinfo {volume}
  {103}},\ \bibinfo {pages} {124067} (\bibinfo {year} {2021}{\natexlab{a}})},\
  \Eprint {http://arxiv.org/abs/2103.08555} {arXiv:2103.08555 [gr-qc]}
  \BibitemShut {NoStop}%
\bibitem [{\citenamefont {Kol\'a\v{r}}\ \emph
  {et~al.}(2021{\natexlab{b}})\citenamefont {Kol\'a\v{r}}, \citenamefont
  {M\'alek}, \citenamefont {Dengiz},\ and\ \citenamefont
  {Kilicarslan}}]{Kolar:2021uiu}%
  \BibitemOpen
  \bibfield  {author} {\bibinfo {author} {\bibfnamefont {I.}~\bibnamefont
  {Kol\'a\v{r}}}, \bibinfo {author} {\bibfnamefont {T.}~\bibnamefont
  {M\'alek}}, \bibinfo {author} {\bibfnamefont {S.}~\bibnamefont {Dengiz}}, \
  and\ \bibinfo {author} {\bibfnamefont {E.}~\bibnamefont {Kilicarslan}},\
  }\href@noop {} {\  (\bibinfo {year} {2021}{\natexlab{b}})},\ \Eprint
  {http://arxiv.org/abs/2107.11884} {arXiv:2107.11884 [gr-qc]} \BibitemShut
  {NoStop}%
\bibitem [{\citenamefont {Llosa}\ and\ \citenamefont
  {Vives}(1994)}]{llosa1994hamiltonian}%
  \BibitemOpen
  \bibfield  {author} {\bibinfo {author} {\bibfnamefont {J.}~\bibnamefont
  {Llosa}}\ and\ \bibinfo {author} {\bibfnamefont {J.}~\bibnamefont {Vives}},\
  }\href@noop {} {\bibfield  {journal} {\bibinfo  {journal} {Journal of
  Mathematical Physics}\ }\textbf {\bibinfo {volume} {35}},\ \bibinfo {pages}
  {2856} (\bibinfo {year} {1994})}\BibitemShut {NoStop}%
\bibitem [{\citenamefont {Gomis}\ \emph {et~al.}(2001)\citenamefont {Gomis},
  \citenamefont {Kamimura},\ and\ \citenamefont {Llosa}}]{Gomis:2000gy}%
  \BibitemOpen
  \bibfield  {author} {\bibinfo {author} {\bibfnamefont {J.}~\bibnamefont
  {Gomis}}, \bibinfo {author} {\bibfnamefont {K.}~\bibnamefont {Kamimura}}, \
  and\ \bibinfo {author} {\bibfnamefont {J.}~\bibnamefont {Llosa}},\ }\href
  {\doibase 10.1103/PhysRevD.63.045003} {\bibfield  {journal} {\bibinfo
  {journal} {Phys. Rev. D}\ }\textbf {\bibinfo {volume} {63}},\ \bibinfo
  {pages} {045003} (\bibinfo {year} {2001})},\ \Eprint
  {http://arxiv.org/abs/hep-th/0006235} {arXiv:hep-th/0006235} \BibitemShut
  {NoStop}%
\bibitem [{\citenamefont {Gomis}\ \emph {et~al.}(2004)\citenamefont {Gomis},
  \citenamefont {Kamimura},\ and\ \citenamefont {Ramirez}}]{Gomis:2003xv}%
  \BibitemOpen
  \bibfield  {author} {\bibinfo {author} {\bibfnamefont {J.}~\bibnamefont
  {Gomis}}, \bibinfo {author} {\bibfnamefont {K.}~\bibnamefont {Kamimura}}, \
  and\ \bibinfo {author} {\bibfnamefont {T.}~\bibnamefont {Ramirez}},\ }\href
  {\doibase 10.1016/j.nuclphysb.2004.06.046} {\bibfield  {journal} {\bibinfo
  {journal} {Nucl. Phys. B}\ }\textbf {\bibinfo {volume} {696}},\ \bibinfo
  {pages} {263} (\bibinfo {year} {2004})},\ \Eprint
  {http://arxiv.org/abs/hep-th/0311184} {arXiv:hep-th/0311184} \BibitemShut
  {NoStop}%
\bibitem [{\citenamefont {Kolar}\ and\ \citenamefont
  {Mazumdar}(2020)}]{Kolar:2020ezu}%
  \BibitemOpen
  \bibfield  {author} {\bibinfo {author} {\bibfnamefont {I.}~\bibnamefont
  {Kolar}}\ and\ \bibinfo {author} {\bibfnamefont {A.}~\bibnamefont
  {Mazumdar}},\ }\href {\doibase 10.1103/PhysRevD.101.124028} {\bibfield
  {journal} {\bibinfo  {journal} {Phys. Rev. D}\ }\textbf {\bibinfo {volume}
  {101}},\ \bibinfo {pages} {124028} (\bibinfo {year} {2020})},\ \Eprint
  {http://arxiv.org/abs/2003.00590} {arXiv:2003.00590 [gr-qc]} \BibitemShut
  {NoStop}%
\bibitem [{\citenamefont {Heredia}\ and\ \citenamefont
  {Llosa}(2021)}]{Heredia:2021wja}%
  \BibitemOpen
  \bibfield  {author} {\bibinfo {author} {\bibfnamefont {C.}~\bibnamefont
  {Heredia}}\ and\ \bibinfo {author} {\bibfnamefont {J.}~\bibnamefont
  {Llosa}},\ }\href {\doibase 10.1088/1751-8121/ac265c} {\bibfield  {journal}
  {\bibinfo  {journal} {J. Phys. A}\ }\textbf {\bibinfo {volume} {54}},\
  \bibinfo {pages} {425202} (\bibinfo {year} {2021})},\ \Eprint
  {http://arxiv.org/abs/2105.10442} {arXiv:2105.10442 [hep-th]} \BibitemShut
  {NoStop}%
\bibitem [{\citenamefont {Heredia}\ and\ \citenamefont
  {Llosa}(2022)}]{Heredia:2022mls}%
  \BibitemOpen
  \bibfield  {author} {\bibinfo {author} {\bibfnamefont {C.}~\bibnamefont
  {Heredia}}\ and\ \bibinfo {author} {\bibfnamefont {J.}~\bibnamefont
  {Llosa}},\ }\href@noop {} {\  (\bibinfo {year} {2022})},\ \Eprint
  {http://arxiv.org/abs/2203.02206} {arXiv:2203.02206 [hep-th]} \BibitemShut
  {NoStop}%
\bibitem [{\citenamefont {Modesto}\ \emph {et~al.}(2018)\citenamefont
  {Modesto}, \citenamefont {Rachwa\l{}},\ and\ \citenamefont
  {Shapiro}}]{Modesto:2017hzl}%
  \BibitemOpen
  \bibfield  {author} {\bibinfo {author} {\bibfnamefont {L.}~\bibnamefont
  {Modesto}}, \bibinfo {author} {\bibfnamefont {L.}~\bibnamefont {Rachwa\l{}}},
  \ and\ \bibinfo {author} {\bibfnamefont {I.~L.}\ \bibnamefont {Shapiro}},\
  }\href {\doibase 10.1140/epjc/s10052-018-6035-2} {\bibfield  {journal}
  {\bibinfo  {journal} {Eur. Phys. J. C}\ }\textbf {\bibinfo {volume} {78}},\
  \bibinfo {pages} {555} (\bibinfo {year} {2018})},\ \Eprint
  {http://arxiv.org/abs/1704.03988} {arXiv:1704.03988 [hep-th]} \BibitemShut
  {NoStop}%
\bibitem [{\citenamefont {Modesto}\ and\ \citenamefont
  {Rachwal}(2014)}]{Modesto:2014lga}%
  \BibitemOpen
  \bibfield  {author} {\bibinfo {author} {\bibfnamefont {L.}~\bibnamefont
  {Modesto}}\ and\ \bibinfo {author} {\bibfnamefont {L.}~\bibnamefont
  {Rachwal}},\ }\href {\doibase 10.1016/j.nuclphysb.2014.10.015} {\bibfield
  {journal} {\bibinfo  {journal} {Nucl. Phys. B}\ }\textbf {\bibinfo {volume}
  {889}},\ \bibinfo {pages} {228} (\bibinfo {year} {2014})},\ \Eprint
  {http://arxiv.org/abs/1407.8036} {arXiv:1407.8036 [hep-th]} \BibitemShut
  {NoStop}%
\bibitem [{\citenamefont {Boos}\ and\ \citenamefont
  {Carone}(2022)}]{Boos:2021lsj}%
  \BibitemOpen
  \bibfield  {author} {\bibinfo {author} {\bibfnamefont {J.}~\bibnamefont
  {Boos}}\ and\ \bibinfo {author} {\bibfnamefont {C.~D.}\ \bibnamefont
  {Carone}},\ }\href {\doibase 10.1103/PhysRevD.105.035034} {\bibfield
  {journal} {\bibinfo  {journal} {Phys. Rev. D}\ }\textbf {\bibinfo {volume}
  {105}},\ \bibinfo {pages} {035034} (\bibinfo {year} {2022})},\ \Eprint
  {http://arxiv.org/abs/2112.05270} {arXiv:2112.05270 [hep-ph]} \BibitemShut
  {NoStop}%
\bibitem [{\citenamefont {Boos}\ and\ \citenamefont
  {Carone}(2021{\natexlab{a}})}]{Boos:2021jih}%
  \BibitemOpen
  \bibfield  {author} {\bibinfo {author} {\bibfnamefont {J.}~\bibnamefont
  {Boos}}\ and\ \bibinfo {author} {\bibfnamefont {C.~D.}\ \bibnamefont
  {Carone}},\ }\href {\doibase 10.1103/PhysRevD.104.095020} {\bibfield
  {journal} {\bibinfo  {journal} {Phys. Rev. D}\ }\textbf {\bibinfo {volume}
  {104}},\ \bibinfo {pages} {095020} (\bibinfo {year} {2021}{\natexlab{a}})},\
  \Eprint {http://arxiv.org/abs/2109.06261} {arXiv:2109.06261 [hep-th]}
  \BibitemShut {NoStop}%
\bibitem [{\citenamefont {Boos}\ and\ \citenamefont
  {Carone}(2021{\natexlab{b}})}]{Boos:2021chb}%
  \BibitemOpen
  \bibfield  {author} {\bibinfo {author} {\bibfnamefont {J.}~\bibnamefont
  {Boos}}\ and\ \bibinfo {author} {\bibfnamefont {C.~D.}\ \bibnamefont
  {Carone}},\ }\href {\doibase 10.1103/PhysRevD.104.015028} {\bibfield
  {journal} {\bibinfo  {journal} {Phys. Rev. D}\ }\textbf {\bibinfo {volume}
  {104}},\ \bibinfo {pages} {015028} (\bibinfo {year} {2021}{\natexlab{b}})},\
  \Eprint {http://arxiv.org/abs/2104.11195} {arXiv:2104.11195 [hep-th]}
  \BibitemShut {NoStop}%
\bibitem [{\citenamefont {Bose}\ \emph {et~al.}(2023)\citenamefont {Bose},
  \citenamefont {Mazumdar}, \citenamefont {Schut},\ and\ \citenamefont
  {Toro\v{s}}}]{Bose:2022czr}%
  \BibitemOpen
  \bibfield  {author} {\bibinfo {author} {\bibfnamefont {S.}~\bibnamefont
  {Bose}}, \bibinfo {author} {\bibfnamefont {A.}~\bibnamefont {Mazumdar}},
  \bibinfo {author} {\bibfnamefont {M.}~\bibnamefont {Schut}}, \ and\ \bibinfo
  {author} {\bibfnamefont {M.}~\bibnamefont {Toro\v{s}}},\ }\href {\doibase
  10.3390/e25030448} {\bibfield  {journal} {\bibinfo  {journal} {Entropy}\
  }\textbf {\bibinfo {volume} {25}},\ \bibinfo {pages} {448} (\bibinfo {year}
  {2023})},\ \Eprint {http://arxiv.org/abs/2203.11628} {arXiv:2203.11628
  [gr-qc]} \BibitemShut {NoStop}%
\bibitem [{\citenamefont {Elahi}\ and\ \citenamefont
  {Mazumdar}(2023)}]{Elahi:2023ozf}%
  \BibitemOpen
  \bibfield  {author} {\bibinfo {author} {\bibfnamefont {S.~G.}\ \bibnamefont
  {Elahi}}\ and\ \bibinfo {author} {\bibfnamefont {A.}~\bibnamefont
  {Mazumdar}},\ }\href@noop {} {\  (\bibinfo {year} {2023})},\ \Eprint
  {http://arxiv.org/abs/2303.07371} {arXiv:2303.07371 [gr-qc]} \BibitemShut
  {NoStop}%
\bibitem [{\citenamefont {Gupta}(1952)}]{Gupta}%
  \BibitemOpen
  \bibfield  {author} {\bibinfo {author} {\bibfnamefont {S.~N.}\ \bibnamefont
  {Gupta}},\ }\href {\doibase 10.1088/0370-1298/65/3/301} {\bibfield  {journal}
  {\bibinfo  {journal} {Proceedings of the Physical Society. Section A}\
  }\textbf {\bibinfo {volume} {65}},\ \bibinfo {pages} {161} (\bibinfo {year}
  {1952})}\BibitemShut {NoStop}%
\bibitem [{\citenamefont {Nguyen}\ and\ \citenamefont
  {Bernards}(2019)}]{Nguyen:2019huk}%
  \BibitemOpen
  \bibfield  {author} {\bibinfo {author} {\bibfnamefont {H.~C.}\ \bibnamefont
  {Nguyen}}\ and\ \bibinfo {author} {\bibfnamefont {F.}~\bibnamefont
  {Bernards}},\ }\href@noop {} {\  (\bibinfo {year} {2019})},\ \Eprint
  {http://arxiv.org/abs/1906.11184} {arXiv:1906.11184 [quant-ph]} \BibitemShut
  {NoStop}%
\bibitem [{\citenamefont {Einstein}(1918)}]{Einstein:1918btx}%
  \BibitemOpen
  \bibfield  {author} {\bibinfo {author} {\bibfnamefont {A.}~\bibnamefont
  {Einstein}},\ }\href@noop {} {\bibfield  {journal} {\bibinfo  {journal}
  {Sitzungsber. Preuss. Akad. Wiss. Berlin (Math. Phys. )}\ }\textbf {\bibinfo
  {volume} {1918}},\ \bibinfo {pages} {154} (\bibinfo {year}
  {1918})}\BibitemShut {NoStop}%
\bibitem [{\citenamefont {Schwinger}(1948)}]{Schwinger:1948yk}%
  \BibitemOpen
  \bibfield  {author} {\bibinfo {author} {\bibfnamefont {J.~S.}\ \bibnamefont
  {Schwinger}},\ }\href {\doibase 10.1103/PhysRev.74.1439} {\bibfield
  {journal} {\bibinfo  {journal} {Phys. Rev.}\ }\textbf {\bibinfo {volume}
  {74}},\ \bibinfo {pages} {1439} (\bibinfo {year} {1948})}\BibitemShut
  {NoStop}%
\bibitem [{\citenamefont {Gupta}(1950)}]{Suraj_N_Gupta_1950}%
  \BibitemOpen
  \bibfield  {author} {\bibinfo {author} {\bibfnamefont {S.~N.}\ \bibnamefont
  {Gupta}},\ }\href {\doibase 10.1088/0370-1298/63/7/301} {\bibfield  {journal}
  {\bibinfo  {journal} {Proceedings of the Physical Society. Section A}\
  }\textbf {\bibinfo {volume} {63}},\ \bibinfo {pages} {681} (\bibinfo {year}
  {1950})}\BibitemShut {NoStop}%
\bibitem [{\citenamefont {Weyl}(1927)}]{Weyl:1927vd}%
  \BibitemOpen
  \bibfield  {author} {\bibinfo {author} {\bibfnamefont {H.}~\bibnamefont
  {Weyl}},\ }\href {\doibase 10.1007/BF02055756} {\bibfield  {journal}
  {\bibinfo  {journal} {Z. Phys.}\ }\textbf {\bibinfo {volume} {46}},\ \bibinfo
  {pages} {1} (\bibinfo {year} {1927})}\BibitemShut {NoStop}%
\bibitem [{\citenamefont {Scadron}(1979)}]{michael1979advanced}%
  \BibitemOpen
  \bibfield  {author} {\bibinfo {author} {\bibfnamefont {M.~D.}\ \bibnamefont
  {Scadron}},\ }\href@noop {} {\emph {\bibinfo {title} {Advanced Quantum Theory
  and Its Applications Through Feynman Diagrams}}}\ (\bibinfo  {publisher}
  {Springer-Verlag.},\ \bibinfo {year} {1979})\BibitemShut {NoStop}%
\bibitem [{\citenamefont {Sakurai}\ and\ \citenamefont
  {Napolitano}(2017)}]{sakurai}%
  \BibitemOpen
  \bibfield  {author} {\bibinfo {author} {\bibfnamefont {J.}~\bibnamefont
  {Sakurai}}\ and\ \bibinfo {author} {\bibfnamefont {J.}~\bibnamefont
  {Napolitano}},\ }\href {https://books.google.co.za/books?id=010yDwAAQBAJ}
  {\emph {\bibinfo {title} {Modern Quantum Mechanics}}},\ \bibinfo {edition}
  {2nd}\ ed.\ (\bibinfo  {publisher} {Cambridge University Press},\ \bibinfo
  {address} {Cambridge U.K.},\ \bibinfo {year} {2017})\BibitemShut {NoStop}%
\bibitem [{\citenamefont {Rungta}\ \emph {et~al.}(2001)\citenamefont {Rungta},
  \citenamefont {Bu{\v{z} }ek}, \citenamefont {Caves}, \citenamefont
  {Hillery},\ and\ \citenamefont {Milburn}}]{rungta}%
  \BibitemOpen
  \bibfield  {author} {\bibinfo {author} {\bibfnamefont {P.}~\bibnamefont
  {Rungta}}, \bibinfo {author} {\bibfnamefont {V.}~\bibnamefont {Bu{\v{z}
  }ek}}, \bibinfo {author} {\bibfnamefont {C.~M.}\ \bibnamefont {Caves}},
  \bibinfo {author} {\bibfnamefont {M.}~\bibnamefont {Hillery}}, \ and\
  \bibinfo {author} {\bibfnamefont {G.~J.}\ \bibnamefont {Milburn}},\ }\href
  {\doibase 10.1103/physreva.64.042315} {\bibfield  {journal} {\bibinfo
  {journal} {Physical Review A}\ }\textbf {\bibinfo {volume} {64}} (\bibinfo
  {year} {2001}),\ 10.1103/physreva.64.042315}\BibitemShut {NoStop}%
\bibitem [{\citenamefont {Casimir}\ and\ \citenamefont
  {Polder}(1948)}]{Casimir:1947kzi}%
  \BibitemOpen
  \bibfield  {author} {\bibinfo {author} {\bibfnamefont {H.~B.~G.}\
  \bibnamefont {Casimir}}\ and\ \bibinfo {author} {\bibfnamefont
  {D.}~\bibnamefont {Polder}},\ }\href {\doibase 10.1103/PhysRev.73.360}
  {\bibfield  {journal} {\bibinfo  {journal} {Phys. Rev.}\ }\textbf {\bibinfo
  {volume} {73}},\ \bibinfo {pages} {360} (\bibinfo {year} {1948})}\BibitemShut
  {NoStop}%
\bibitem [{\citenamefont {Casimir}(1948)}]{Casimir:1948dh}%
  \BibitemOpen
  \bibfield  {author} {\bibinfo {author} {\bibfnamefont {H.~B.~G.}\
  \bibnamefont {Casimir}},\ }\href@noop {} {\bibfield  {journal} {\bibinfo
  {journal} {Indag. Math.}\ }\textbf {\bibinfo {volume} {10}},\ \bibinfo
  {pages} {261} (\bibinfo {year} {1948})}\BibitemShut {NoStop}%
\bibitem [{\citenamefont {Barker}\ \emph {et~al.}(2022)\citenamefont {Barker},
  \citenamefont {Bose}, \citenamefont {Marshman},\ and\ \citenamefont
  {Mazumdar}}]{Barker:2022mdz}%
  \BibitemOpen
  \bibfield  {author} {\bibinfo {author} {\bibfnamefont {P.~F.}\ \bibnamefont
  {Barker}}, \bibinfo {author} {\bibfnamefont {S.}~\bibnamefont {Bose}},
  \bibinfo {author} {\bibfnamefont {R.~J.}\ \bibnamefont {Marshman}}, \ and\
  \bibinfo {author} {\bibfnamefont {A.}~\bibnamefont {Mazumdar}},\ }\href
  {\doibase 10.1103/PhysRevD.106.L041901} {\bibfield  {journal} {\bibinfo
  {journal} {Phys. Rev. D}\ }\textbf {\bibinfo {volume} {106}},\ \bibinfo
  {pages} {L041901} (\bibinfo {year} {2022})},\ \Eprint
  {http://arxiv.org/abs/2203.00038} {arXiv:2203.00038 [hep-ph]} \BibitemShut
  {NoStop}%
\bibitem [{\citenamefont {Nielsen}\ and\ \citenamefont
  {Chuang}(2010)}]{nielsen_chuang_2010}%
  \BibitemOpen
  \bibfield  {author} {\bibinfo {author} {\bibfnamefont {M.~A.}\ \bibnamefont
  {Nielsen}}\ and\ \bibinfo {author} {\bibfnamefont {I.~L.}\ \bibnamefont
  {Chuang}},\ }\href {\doibase 10.1017/CBO9780511976667} {\emph {\bibinfo
  {title} {Quantum Computation and Quantum Information: 10th Anniversary
  Edition}}}\ (\bibinfo  {publisher} {Cambridge University Press},\ \bibinfo
  {year} {2010})\BibitemShut {NoStop}%
\bibitem [{\citenamefont {Tilly}\ \emph {et~al.}(2021)\citenamefont {Tilly},
  \citenamefont {Marshman}, \citenamefont {Mazumdar},\ and\ \citenamefont
  {Bose}}]{Tilly:2021qef}%
  \BibitemOpen
  \bibfield  {author} {\bibinfo {author} {\bibfnamefont {J.}~\bibnamefont
  {Tilly}}, \bibinfo {author} {\bibfnamefont {R.~J.}\ \bibnamefont {Marshman}},
  \bibinfo {author} {\bibfnamefont {A.}~\bibnamefont {Mazumdar}}, \ and\
  \bibinfo {author} {\bibfnamefont {S.}~\bibnamefont {Bose}},\ }\href {\doibase
  10.1103/PhysRevA.104.052416} {\bibfield  {journal} {\bibinfo  {journal}
  {Phys. Rev. A}\ }\textbf {\bibinfo {volume} {104}},\ \bibinfo {pages}
  {052416} (\bibinfo {year} {2021})},\ \Eprint
  {http://arxiv.org/abs/2101.08086} {arXiv:2101.08086 [quant-ph]} \BibitemShut
  {NoStop}%
\bibitem [{\citenamefont {Schut}\ \emph {et~al.}(2022)\citenamefont {Schut},
  \citenamefont {Tilly}, \citenamefont {Marshman}, \citenamefont {Bose},\ and\
  \citenamefont {Mazumdar}}]{Schut:2021svd}%
  \BibitemOpen
  \bibfield  {author} {\bibinfo {author} {\bibfnamefont {M.}~\bibnamefont
  {Schut}}, \bibinfo {author} {\bibfnamefont {J.}~\bibnamefont {Tilly}},
  \bibinfo {author} {\bibfnamefont {R.~J.}\ \bibnamefont {Marshman}}, \bibinfo
  {author} {\bibfnamefont {S.}~\bibnamefont {Bose}}, \ and\ \bibinfo {author}
  {\bibfnamefont {A.}~\bibnamefont {Mazumdar}},\ }\href {\doibase
  10.1103/PhysRevA.105.032411} {\bibfield  {journal} {\bibinfo  {journal}
  {Phys. Rev. A}\ }\textbf {\bibinfo {volume} {105}},\ \bibinfo {pages}
  {032411} (\bibinfo {year} {2022})},\ \Eprint
  {http://arxiv.org/abs/2110.14695} {arXiv:2110.14695 [quant-ph]} \BibitemShut
  {NoStop}%
\bibitem [{\citenamefont {Lee}\ \emph {et~al.}(2020)\citenamefont {Lee},
  \citenamefont {Adelberger}, \citenamefont {Cook}, \citenamefont {Fleischer},\
  and\ \citenamefont {Heckel}}]{Lee_2020}%
  \BibitemOpen
  \bibfield  {author} {\bibinfo {author} {\bibfnamefont {J.}~\bibnamefont
  {Lee}}, \bibinfo {author} {\bibfnamefont {E.}~\bibnamefont {Adelberger}},
  \bibinfo {author} {\bibfnamefont {T.}~\bibnamefont {Cook}}, \bibinfo {author}
  {\bibfnamefont {S.}~\bibnamefont {Fleischer}}, \ and\ \bibinfo {author}
  {\bibfnamefont {B.}~\bibnamefont {Heckel}},\ }\href {\doibase
  10.1103/physrevlett.124.101101} {\bibfield  {journal} {\bibinfo  {journal}
  {Physical Review Letters}\ }\textbf {\bibinfo {volume} {124}} (\bibinfo
  {year} {2020}),\ 10.1103/physrevlett.124.101101}\BibitemShut {NoStop}%
\bibitem [{\citenamefont {Marshman}\ \emph {et~al.}(2022)\citenamefont
  {Marshman}, \citenamefont {Mazumdar}, \citenamefont {Folman},\ and\
  \citenamefont {Bose}}]{Marshman:2021wyk}%
  \BibitemOpen
  \bibfield  {author} {\bibinfo {author} {\bibfnamefont {R.~J.}\ \bibnamefont
  {Marshman}}, \bibinfo {author} {\bibfnamefont {A.}~\bibnamefont {Mazumdar}},
  \bibinfo {author} {\bibfnamefont {R.}~\bibnamefont {Folman}}, \ and\ \bibinfo
  {author} {\bibfnamefont {S.}~\bibnamefont {Bose}},\ }\href {\doibase
  10.1103/PhysRevResearch.4.023087} {\bibfield  {journal} {\bibinfo  {journal}
  {Phys. Rev. Res.}\ }\textbf {\bibinfo {volume} {4}},\ \bibinfo {pages}
  {023087} (\bibinfo {year} {2022})},\ \Eprint
  {http://arxiv.org/abs/2105.01094} {arXiv:2105.01094 [quant-ph]} \BibitemShut
  {NoStop}%
\bibitem [{\citenamefont {Pedernales}\ \emph {et~al.}(2020)\citenamefont
  {Pedernales}, \citenamefont {Morley},\ and\ \citenamefont
  {Plenio}}]{Pedernales2020MotionalDD}%
  \BibitemOpen
  \bibfield  {author} {\bibinfo {author} {\bibfnamefont {J.~S.}\ \bibnamefont
  {Pedernales}}, \bibinfo {author} {\bibfnamefont {G.~W.}\ \bibnamefont
  {Morley}}, \ and\ \bibinfo {author} {\bibfnamefont {M.~B.}\ \bibnamefont
  {Plenio}},\ }\href@noop {} {\bibfield  {journal} {\bibinfo  {journal}
  {Physical review letters}\ }\textbf {\bibinfo {volume} {125 2}},\ \bibinfo
  {pages} {023602} (\bibinfo {year} {2020})}\BibitemShut {NoStop}%
\bibitem [{\citenamefont {Zhou}\ \emph
  {et~al.}(2022{\natexlab{a}})\citenamefont {Zhou}, \citenamefont {Marshman},
  \citenamefont {Bose},\ and\ \citenamefont {Mazumdar}}]{Zhou:2022frl}%
  \BibitemOpen
  \bibfield  {author} {\bibinfo {author} {\bibfnamefont {R.}~\bibnamefont
  {Zhou}}, \bibinfo {author} {\bibfnamefont {R.~J.}\ \bibnamefont {Marshman}},
  \bibinfo {author} {\bibfnamefont {S.}~\bibnamefont {Bose}}, \ and\ \bibinfo
  {author} {\bibfnamefont {A.}~\bibnamefont {Mazumdar}},\ }\href {\doibase
  10.1103/PhysRevResearch.4.043157} {\bibfield  {journal} {\bibinfo  {journal}
  {Phys. Rev. Res.}\ }\textbf {\bibinfo {volume} {4}},\ \bibinfo {pages}
  {043157} (\bibinfo {year} {2022}{\natexlab{a}})},\ \Eprint
  {http://arxiv.org/abs/2206.04088} {arXiv:2206.04088 [quant-ph]} \BibitemShut
  {NoStop}%
\bibitem [{\citenamefont {Zhou}\ \emph {et~al.}(2023)\citenamefont {Zhou},
  \citenamefont {Marshman}, \citenamefont {Bose},\ and\ \citenamefont
  {Mazumdar}}]{Zhou:2022jug}%
  \BibitemOpen
  \bibfield  {author} {\bibinfo {author} {\bibfnamefont {R.}~\bibnamefont
  {Zhou}}, \bibinfo {author} {\bibfnamefont {R.~J.}\ \bibnamefont {Marshman}},
  \bibinfo {author} {\bibfnamefont {S.}~\bibnamefont {Bose}}, \ and\ \bibinfo
  {author} {\bibfnamefont {A.}~\bibnamefont {Mazumdar}},\ }\href {\doibase
  10.1103/PhysRevA.107.032212} {\bibfield  {journal} {\bibinfo  {journal}
  {Phys. Rev. A}\ }\textbf {\bibinfo {volume} {107}},\ \bibinfo {pages}
  {032212} (\bibinfo {year} {2023})},\ \Eprint
  {http://arxiv.org/abs/2210.05689} {arXiv:2210.05689 [quant-ph]} \BibitemShut
  {NoStop}%
\bibitem [{\citenamefont {Zhou}\ \emph
  {et~al.}(2022{\natexlab{b}})\citenamefont {Zhou}, \citenamefont {Marshman},
  \citenamefont {Bose},\ and\ \citenamefont {Mazumdar}}]{Zhou:2022epb}%
  \BibitemOpen
  \bibfield  {author} {\bibinfo {author} {\bibfnamefont {R.}~\bibnamefont
  {Zhou}}, \bibinfo {author} {\bibfnamefont {R.~J.}\ \bibnamefont {Marshman}},
  \bibinfo {author} {\bibfnamefont {S.}~\bibnamefont {Bose}}, \ and\ \bibinfo
  {author} {\bibfnamefont {A.}~\bibnamefont {Mazumdar}},\ }\href@noop {} {\
  (\bibinfo {year} {2022}{\natexlab{b}})},\ \Eprint
  {http://arxiv.org/abs/2211.08435} {arXiv:2211.08435 [quant-ph]} \BibitemShut
  {NoStop}%
\bibitem [{\citenamefont {Toro\v{s}}\ \emph {et~al.}(2021)\citenamefont
  {Toro\v{s}}, \citenamefont {Van De~Kamp}, \citenamefont {Marshman},
  \citenamefont {Kim}, \citenamefont {Mazumdar},\ and\ \citenamefont
  {Bose}}]{Toros:2020dbf}%
  \BibitemOpen
  \bibfield  {author} {\bibinfo {author} {\bibfnamefont {M.}~\bibnamefont
  {Toro\v{s}}}, \bibinfo {author} {\bibfnamefont {T.~W.}\ \bibnamefont {Van
  De~Kamp}}, \bibinfo {author} {\bibfnamefont {R.~J.}\ \bibnamefont
  {Marshman}}, \bibinfo {author} {\bibfnamefont {M.~S.}\ \bibnamefont {Kim}},
  \bibinfo {author} {\bibfnamefont {A.}~\bibnamefont {Mazumdar}}, \ and\
  \bibinfo {author} {\bibfnamefont {S.}~\bibnamefont {Bose}},\ }\href {\doibase
  10.1103/PhysRevResearch.3.023178} {\bibfield  {journal} {\bibinfo  {journal}
  {Phys. Rev. Res.}\ }\textbf {\bibinfo {volume} {3}},\ \bibinfo {pages}
  {023178} (\bibinfo {year} {2021})},\ \Eprint
  {http://arxiv.org/abs/2007.15029} {arXiv:2007.15029 [gr-qc]} \BibitemShut
  {NoStop}%
\bibitem [{\citenamefont {Gunnink}\ \emph {et~al.}(2022)\citenamefont
  {Gunnink}, \citenamefont {Mazumdar}, \citenamefont {Schut},\ and\
  \citenamefont {Toro\v{s}}}]{Gunnink:2022ner}%
  \BibitemOpen
  \bibfield  {author} {\bibinfo {author} {\bibfnamefont {F.}~\bibnamefont
  {Gunnink}}, \bibinfo {author} {\bibfnamefont {A.}~\bibnamefont {Mazumdar}},
  \bibinfo {author} {\bibfnamefont {M.}~\bibnamefont {Schut}}, \ and\ \bibinfo
  {author} {\bibfnamefont {M.}~\bibnamefont {Toro\v{s}}},\ }\href@noop {} {\
  (\bibinfo {year} {2022})},\ \Eprint {http://arxiv.org/abs/2210.16919}
  {arXiv:2210.16919 [quant-ph]} \BibitemShut {NoStop}%
\bibitem [{\citenamefont {Wu}\ \emph {et~al.}(2022)\citenamefont {Wu},
  \citenamefont {Toro\v{s}}, \citenamefont {Bose},\ and\ \citenamefont
  {Mazumdar}}]{Wu:2022rdv}%
  \BibitemOpen
  \bibfield  {author} {\bibinfo {author} {\bibfnamefont {M.-Z.}\ \bibnamefont
  {Wu}}, \bibinfo {author} {\bibfnamefont {M.}~\bibnamefont {Toro\v{s}}},
  \bibinfo {author} {\bibfnamefont {S.}~\bibnamefont {Bose}}, \ and\ \bibinfo
  {author} {\bibfnamefont {A.}~\bibnamefont {Mazumdar}},\ }\href@noop {} {\
  (\bibinfo {year} {2022})},\ \Eprint {http://arxiv.org/abs/2211.15695}
  {arXiv:2211.15695 [gr-qc]} \BibitemShut {NoStop}%
\bibitem [{\citenamefont {van~de Kamp}\ \emph {et~al.}(2020)\citenamefont
  {van~de Kamp}, \citenamefont {Marshman}, \citenamefont {Bose},\ and\
  \citenamefont {Mazumdar}}]{vandeKamp:2020rqh}%
  \BibitemOpen
  \bibfield  {author} {\bibinfo {author} {\bibfnamefont {T.~W.}\ \bibnamefont
  {van~de Kamp}}, \bibinfo {author} {\bibfnamefont {R.~J.}\ \bibnamefont
  {Marshman}}, \bibinfo {author} {\bibfnamefont {S.}~\bibnamefont {Bose}}, \
  and\ \bibinfo {author} {\bibfnamefont {A.}~\bibnamefont {Mazumdar}},\ }\href
  {\doibase 10.1103/PhysRevA.102.062807} {\bibfield  {journal} {\bibinfo
  {journal} {Phys. Rev. A}\ }\textbf {\bibinfo {volume} {102}},\ \bibinfo
  {pages} {062807} (\bibinfo {year} {2020})},\ \Eprint
  {http://arxiv.org/abs/2006.06931} {arXiv:2006.06931 [quant-ph]} \BibitemShut
  {NoStop}%
\bibitem [{\citenamefont {Rijavec}\ \emph {et~al.}(2021)\citenamefont
  {Rijavec}, \citenamefont {Carlesso}, \citenamefont {Bassi}, \citenamefont
  {Vedral},\ and\ \citenamefont {Marletto}}]{Rijavec:2020qxd}%
  \BibitemOpen
  \bibfield  {author} {\bibinfo {author} {\bibfnamefont {S.}~\bibnamefont
  {Rijavec}}, \bibinfo {author} {\bibfnamefont {M.}~\bibnamefont {Carlesso}},
  \bibinfo {author} {\bibfnamefont {A.}~\bibnamefont {Bassi}}, \bibinfo
  {author} {\bibfnamefont {V.}~\bibnamefont {Vedral}}, \ and\ \bibinfo {author}
  {\bibfnamefont {C.}~\bibnamefont {Marletto}},\ }\href {\doibase
  10.1088/1367-2630/abf3eb} {\bibfield  {journal} {\bibinfo  {journal} {New J.
  Phys.}\ }\textbf {\bibinfo {volume} {23}},\ \bibinfo {pages} {043040}
  (\bibinfo {year} {2021})},\ \Eprint {http://arxiv.org/abs/2012.06230}
  {arXiv:2012.06230 [quant-ph]} \BibitemShut {NoStop}%
\bibitem [{\citenamefont {Romero-Isart}\ \emph {et~al.}(2011)\citenamefont
  {Romero-Isart}, \citenamefont {Pflanzer}, \citenamefont {Blaser},
  \citenamefont {Kaltenbaek}, \citenamefont {Kiesel}, \citenamefont
  {Aspelmeyer},\ and\ \citenamefont {Cirac}}]{romero2011large}%
  \BibitemOpen
  \bibfield  {author} {\bibinfo {author} {\bibfnamefont {O.}~\bibnamefont
  {Romero-Isart}}, \bibinfo {author} {\bibfnamefont {A.~C.}\ \bibnamefont
  {Pflanzer}}, \bibinfo {author} {\bibfnamefont {F.}~\bibnamefont {Blaser}},
  \bibinfo {author} {\bibfnamefont {R.}~\bibnamefont {Kaltenbaek}}, \bibinfo
  {author} {\bibfnamefont {N.}~\bibnamefont {Kiesel}}, \bibinfo {author}
  {\bibfnamefont {M.}~\bibnamefont {Aspelmeyer}}, \ and\ \bibinfo {author}
  {\bibfnamefont {J.~I.}\ \bibnamefont {Cirac}},\ }\href {\doibase
  10.1103/PhysRevLett.107.020405} {\bibfield  {journal} {\bibinfo  {journal}
  {Physical Review Letters}\ }\textbf {\bibinfo {volume} {107}},\ \bibinfo
  {pages} {020405} (\bibinfo {year} {2011})}\BibitemShut {NoStop}%
\end{thebibliography}%
\end{document}